\documentclass[authorversion]{acmart} 

\newboolean{anonymized}
\setboolean{anonymized}{false} 

\usepackage{multirow,array}
\usepackage{rotating}

\usepackage{mathdots}
\usepackage{amsthm,amsmath}
\usepackage{mathtools}
\usepackage[ruled, noend]{algorithm2e}
\usepackage{graphicx}
\usepackage{lscape}

\usepackage{cleveref}
\usepackage{enumitem}

\crefname{algorithm}{algorithm}{algorithms}
\Crefname{algorithm}{Algorithm}{Algorithms}

\crefname{line}{algorithm}{algorithms}
\Crefname{line}{Algorithm}{Algorithms}

\AtBeginDocument{%
  \providecommand\BibTeX{{%
    \normalfont B\kern-0.5em{\scshape i\kern-0.25em b}\kern-0.8em\TeX}}}

\setcopyright{acmcopyright}
\copyrightyear{2025}
\acmYear{2025}
\acmDOI{XXXXXXX.XXXXXXX}

\acmPrice{15.00}
\acmISBN{978-1-4503-XXXX-X/18/06}

\begin{document}


\title{``Think about it like you're a firefighter'': Understanding How Reddit Moderators Use the Modqueue}

\author{Tanvi Bajpai}
\email{tbajpai2@illinois.edu}
\affiliation{%
  \institution{University of Illinois Urbana-Champaign}
  \city{Urbana}
  \state{Illinois}
  \country{USA}
  \postcode{61801}
}

\author{Eshwar Chandrasekharan}
\email{eshwar@illinois.edu}
\affiliation{%
  \institution{University of Illinois Urbana-Champaign}
  \city{Urbana}
  \state{Illinois}
  \country{USA}
  \postcode{61801}
}

\renewcommand{\shortauthors}{T. Bajpai and E. Chandrasekharan}

\begin{abstract}

On Reddit, the moderation queue (modqueue) is a primary interface for moderators to review reported content. Despite its central role in Reddit's community-reliant moderation model, little is known about how moderators actually use it in practice. To address this gap, we surveyed 110 moderators, who collectively oversee more than 400 unique subreddits, and asked them about their usage of the modqueue. Modqueue practices vary widely: some moderators approach it as a daily checklist, others as a hub to infer community-wide patterns, and many still find the queue insufficient to inform their moderation decisions. We also identify persistent challenges around review coordination, inconsistent interface signals, and reliance on third-party tools. Taken together, we show the modqueue is neither a one-size-fits-all solution nor sufficient on its own for supporting moderator review. Our work highlights design opportunities for more modular, integrated, and customizable platform infrastructures that better support the diversity of moderator workflows.

\end{abstract}






\maketitle

\section{Introduction}

Platforms like Reddit and Discord utilize community-reliant moderation \cite{caplan2018content}, where volunteer members of the community are deputized to be moderators (\emph{mods}), and are responsible for ensuring that the community remains on task and its members are kept safe. Moderation work in these systems spans a wide range of responsibilities, including rule development and updating \cite{fiesler2018reddit,jiang2020characterizing,seering2019engagement,zhang2020policykit,reddy2023evolution}, configuring automated moderation tools \cite{jhaver19automod,song2023modsandbox,chandrasekharan19crossmod}, and removing rule-violating content or users \cite{jhaver2018online,cai2021after,chandrasekharan2018norms,samory2021positive}. Unlike centralized models of moderation, such as those used on platforms like Facebook or TikTok---which rely primarily on large-scale automated flagging and contracted professionals to enforce platform-wide standards uniformly~\cite{gillespie2018custodians,roberts2019behind,chandrasekharan2017you}---community-reliant models enable norms and community-specific guidelines to be constructed, interpreted, and enforced in ways that reflect the values of each individual community~\cite{chandrasekharan2018norms, goyal2026uncovering}. This nuance and context-sensitivity, however, comes with trade-offs: it places significant responsibility on unpaid volunteer mods , who must balance governance work with everyday participation, and must rely on a combination of platform-provided tools, third-party extensions, and other makeshift solutions to effectively carry out their moderation responsibilities~\cite{kiene2019volunteer, li2022all, li22monetary}.

\subsection{Moderator Review on Reddit}

While mods are responsible for a wide range of governance and maintenance tasks, one of their major responsibilities is to provide oversight on content or users that may be problematic, and subsequently make decisions about whether they are rule-violating (and should thus be removed from the community) or not (in which case they may remain). We will refer to this overarching process as \emph{mod review} for simplicity. 

On Reddit, mod review is prompted when potentially problematic content or users in a community are brought to the attention of the community's mods. This can happen organically when mods encounter issues while browsing their community as ordinary members would. However, relying on organic discovery alone would be impractical, particularly in  large communities where no mod can be aware of all posts or comments at any given moment across discussion threads~\cite{liu2025needling}. 
As such, Reddit provides two\footnote{In reality, there are more than two mechanisms to surface potentially problematic content, since third-party or custom tools are often created by mods and researchers to improve flagging and reporting \cite{reddit2017custombots,reddit2023moderationbots}; we will discuss such tools later on.}  main mechanisms for surfacing such content: (1) a report functionality that allows users to flag content or other users they believe are breaking rules, and (2) the \emph{AutoModerator} (Automod), a rule-based tool that automatically flags or filters using predefined pattern matching \cite{jhaver19automod}. These mechanisms generate a stream of items awaiting mod review. Once flagged, these items appear in the \emph{moderation queue} (modqueue), Reddit’s centralized interface for reviewing reported and filtered content (see \Cref{fig:modq}).

\ifthenelse{\boolean{anonymized}}{
\begin{figure}[!t]
  \centering
  \includegraphics[width=\linewidth]{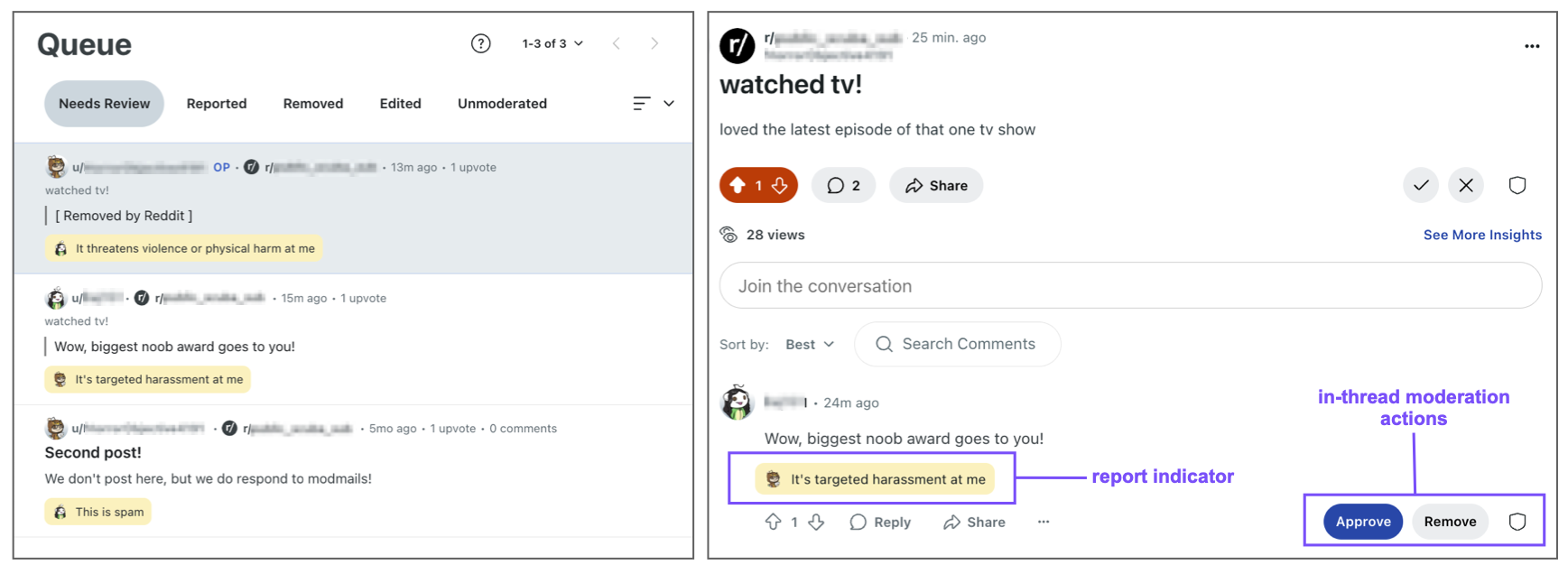}
  \caption{(left) The Reddit moderation queue (modqueue), which centralizes items flagged by users, AutoModerator, or platform filters for moderator review. (right) An example of in-thread moderation functions, allowing mods to view and react directly on a reported post or comment without using the modqueue.}
  \label{fig:modq}
\end{figure}
}{

\begin{figure}[!t]
  \centering
  \includegraphics[width=\linewidth]{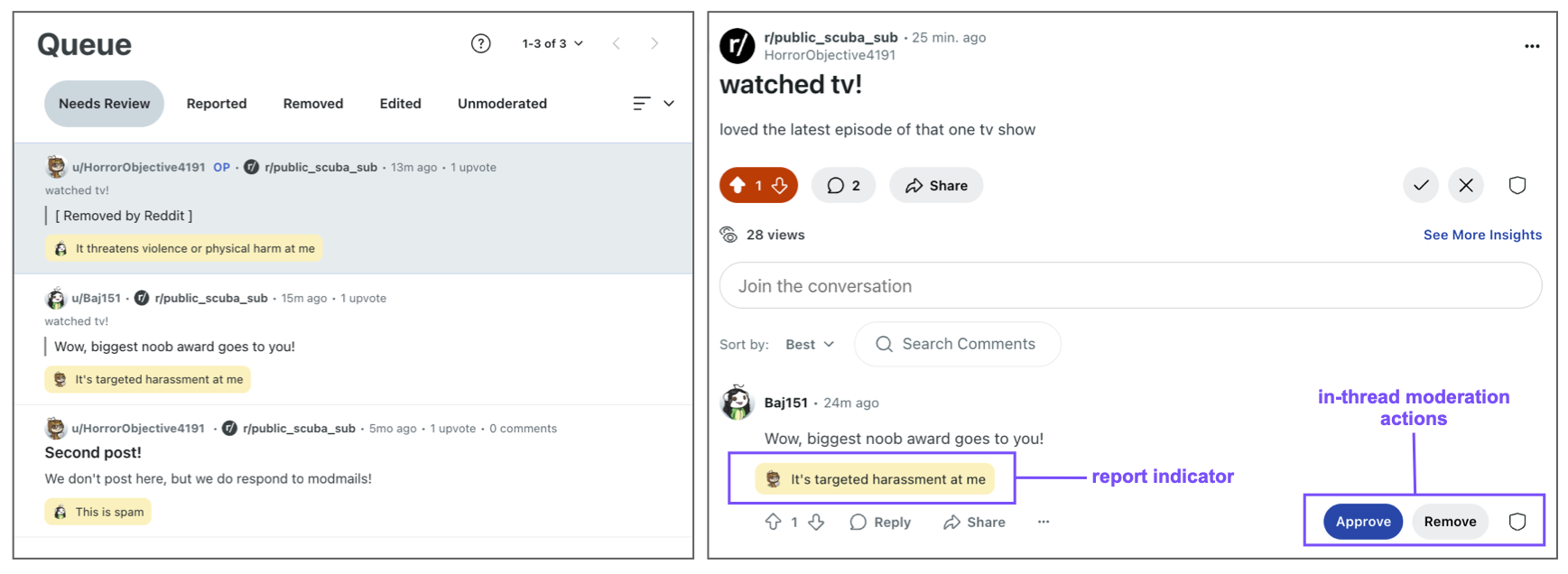}
  \caption{(left) The Reddit moderation queue (modqueue), which centralizes items flagged by users, AutoModerator, or platform filters for moderator review. (right) An example of in-thread moderation functions, allowing mods to view and react directly on a reported post or comment without using the modqueue.}
  \label{fig:modq}
\end{figure}
}

\subsection{Understanding the Realities of Modqueue Usage}

While mods are able to intervene on problematic content directly from within posts or comment threads (see \Cref{fig:modq}), the modqueue is the only interface explicitly designed to triage flagged items across all of the communities they moderate, and the only place on Reddit where Automod-filtered content can be systematically reviewed and validated \cite{modqueue}. As such, the modqueue might be considered a primary venue for mod review on Reddit, and a critical interface that allows mods to reflect the norms and nuances of individual communities.

 Despite its seemingly central role in Reddit’s moderation ecosystem, little research has examined how the modqueue is actually used by mods in practice. Prior moderation research on platforms that utilize a community-reliant approach has largely focused on how communities and mods engage in rule-setting, the design of effective tools and mechanisms for content flagging, user perceptions of moderation outcomes \cite{zhang2020policykit,fiesler2018reddit,jhaver2019did,koshy2023alignment,jhaver19automod,chandrasekharan19crossmod}. In contrast, the realities of mod review practices have received comparatively little attention, and the mod review stage is often treated as a ``black box'' in moderation workflows. This is especially true of the modqueue, which is mentioned only in passing by studies of Reddit moderation \cite{chandrasekharan19crossmod,jhaver19automod,tabassum2024investigating,bozarth2023wisdom}. Understanding how mods use the modqueue in their own moderation workflows is therefore critical to developing a more complete picture of not only the mod review process, but also community-reliant moderation models overall. 

\subsection{Research Questions}

To uncover the realities of modqueue usage, this paper examines how Reddit mods incorporate the modqueue into their moderation workflows. Since prior work has shown that moderation practices on Reddit vary widely across communities and are shaped by differences in size, culture, and goals \cite{chandrasekharan2018norms, fiesler2018reddit}, our goal is to capture this variation and provide a broad understanding of how and why mods engage with the modqueue (or potentially choose not to). More precisely, our aim is to surface a variety of answers to the following research questions:

\begin{itemize}[label={}, leftmargin=10pt]

\item \textbf{(RQ1)} How does the modqueue fit into moderators’ workflows?
\item \textbf{(RQ2)} Do moderators prioritize items within the modqueue, and if so, how?
\item \textbf{(RQ3)} Why would moderators need to leave the modqueue while reviewing items?
\item \textbf{(RQ4)} What challenges do moderators face when using the modqueue?

\end{itemize}

\subsection{Summary of Contributions}

\subsubsection{Methods}
We conducted a survey of 110 active Reddit mods, who collectively moderate over 400 distinct subreddits. Respondents varied widely in moderation experience, subreddit size and topic, and tool/interface configurations. Because Reddit offers multiple interface versions of the modqueue (as outlined in \Cref{sec:modq}), and many mods combine these with third-party extensions or custom tools, our survey questions were deliberately kept high-level and open-ended. This design allowed us to capture a broad spectrum of workflows without assuming a fixed interface or tooling setup. A detailed description of our survey design and methodology is provided in \Cref{sec:methods}.

\subsubsection{Findings}
We find that while most mods use the modqueue, their practices vary widely based on factors like report volume, interface preferences, reliance on automation, and underlying moderation objectives. Some mods treat the modqueue as a daily checklist to clear, while others use it more strategically to surface patterns, monitor community health, or follow up on alerts. Strategies for reviewing reports range from simple sequential traversal to more adaptive or goal-driven approaches (e.g., prioritizing human reports, scanning for high-risk content, or triaging based on urgency, severity, or fairness). Despite Reddit's introduction of new features designed to provide more context within the modqueue interface itself, most mods report regularly leaving the modqueue to investigate surrounding conversation, user history, or past moderation actions taken regarding an item. Finally, we surface several challenges that mods face when using the modqueue: coordination breakdowns within moderation teams during modqueue review work, interface limitations that disrupt multi-step actions, inconsistent or noisy interface signals, and frustration stemming from poor tool integration across Reddit’s multiple interfaces.

\subsubsection{Implications}

The modqueue is not a one-size-fits-all tool: mods adapt it to their personal habits, community needs, and preferred tooling. Platforms like Reddit should therefore support consistent yet modular infrastructures that enable customization and lightweight coordination, better reflecting the collaborative and value-driven nature of community-reliant moderation. The fact that many mods continue to rely on older versions of Reddit’s interface, or selectively ignore new features in updated versions, suggests that platform-driven changes to the modqueue often misalign with mods’ actual workflows. Instead of major interface overhauls, there is a need for stronger integration with the third-party tools that mods already prefer, as well as closer alignment between platform updates and the realities of mod practice. The diversity of modqueue use also points to design opportunities to support not only item-by-item review, but also higher-level pattern recognition across flagged community activity. Specifically, we highlight how the design of future tools could potentially lead to improvements in review efficiency or efficacy, and help in making mod review work more salient and manageable for volunteer mods.

\section{Background}\label{sec:relwork}

To situate our study of how moderators engage with Reddit’s modqueue, we review prior work on community-reliant moderation and on the tools, interfaces, and infrastructures that support moderator practice. We defer background regarding the modqueue itself to \Cref{sec:modq}.

\subsection{Diversity in Community Needs and Moderation Practices}\label{sec:relwork:crm}

Platforms like Reddit, Discord, and Twitch rely on community-reliant models of moderation in part because they host a wide variety of communities with different goals, norms, and practices. More centralized approaches would risk undercutting this diversity by imposing uniform standards that fail to account for community-specific needs. Online community research conducted on these platforms has shown that communities vary in how they establish and enforce rules; these differences are often shaped by factors such as size, topic, and purpose \cite{chandrasekharan2018norms,fiesler2018reddit, reddy2023evolution}. For instance, \citet{fiesler2018reddit} found that while rule categories may overlap across communities on Reddit (subreddits), each subreddit adapts its rules to reflect its own focus and context. Likewise, \citet{chandrasekharan2018norms} demonstrate that norms on Reddit operate at multiple levels: macro-level norms that apply platform-wide, meso-level norms shared across groups of subreddits, and micro-level norms unique to individual subreddits.

Community members’ and moderators’ perspectives also shape how moderation is carried out within communities \cite{gilbert2020run,matias2019civic,weld19values, cai2022coordination}. For example, \citet{gilbert2020run}'s study of r/AskHistorians shows how moderators’ own identities (e.g., seeing themselves as educators, archivists, or stewards \cite{gilbert2020run}) influence the practices they adopt and the perspectives they bring to rule enforcement. In contrast, \citet{weld19values} analyze values articulated by community members across more than 600 subreddits, developing a taxonomy of 29 subcategories of community values. These differences in values can shape how members report content or users they deem inappropriate, antisocial, or rule-violating \cite{jhaver2019did,koshy2023alignment}, which in turn affects the kinds of issues moderators must address and the volume of reports they contend with.

Beyond variation across communities and their members, moderators themselves differ in the types of actions they emphasize when carrying out moderation~\cite{wohn19emotional, cai2023understanding}. Some of these actions are punitive, such as removing rule-violating or abusive content \cite{bruckman1994deviant,gillespie2018custodians,jhaver19automod,chandrasekharan2022quarantined}, while others are constructive, such as encouraging or rewarding pro-social participation and norm-abiding behavior \cite{seering2017pro,seering2019engagement,lambert24positive}. In this sense, moderators working toward the same moderation goals may take very different combinations of actions to accomplish them. Addressing harmful behavior, for instance, might involve removing abusive content, but it could also involve rewarding members who call out problematic posts or model healthier participation. 

Altogether, this body of work highlights the diversity of community moderation needs and the variation in how moderators can address them. For this reason, our study centers on mod review via the modqueue, which serves as a primary venue where diverse community rules and norms are put into practice. Rather than identifying a singular generalized workflow, we aim to capture the wide variety of ways moderators engage with this process.

\subsection{Tools, Interfaces, and Infrastructures for Moderation}\label{sec:relwork:tools}

In addition to the factors discussed in the previous subsection, moderation is heavily shaped by the features, tools, and infrastructures provided by the platforms where it takes place. Prior research has emphasized that the kinds of tools a platform offers directly shape both the practices moderators and users are able to adopt and the challenges they may face \cite{bajpai22mic,jiang19discord, huang2024opportunities}. For instance, \citet{jiang19discord} found that moderators of Discord servers often struggled with issues tied to voice-based infrastructures, such as limited logging, difficulties in collecting evidence, and the ephemeral nature of audio conversations, arguing that technological infrastructure plays a central role in shaping moderation. \citet{bajpai22mic} build on these takeaways by developing a framework that highlights how technological affordances within platform infrastructures make up platforms’ moderation ecosystems, and in turn affect both the work that can be done by moderators and the challenges they encounter; their taxonomy applies not only to platforms that employ community-reliant moderation, but also those that rely on centralized models of moderation.

On platforms that utilize community-reliant moderation, volunteer moderators often work within, around, and against the technological systems provided to them, adapting these infrastructures to meet community needs. On Twitch, chatbots are widely deployed to automate real-time monitoring of livestream chat, filtering out spam or abusive comments or automatically banned, and even highlighting pro-social comments \cite{seering2017pro, cai2019categorizing,seering2023moderates,twitch2025chat}. Discord similarly offers a variety of different applications and chatbots that provide features to support moderation work \cite{discord2025discovery,discord2025modbot}, as well as providing an API that supports the development of custom bots or tools to manage moderation across text and audio channels in individual communities (servers) \cite{jiang19discord, choi23convex, seering2024chillbot, doan2025design,discord2025reference}.

On Reddit specifically, the platform itself provides a range of moderation tools, while third-party developers and moderators extend these capabilities with additional systems~\cite{chang2022thread, koshy2024venire}. For instance, Reddit's AutoModerator tool (Automod) can be configured to automatically flag or remove content based on community-defined patterns \cite{jhaver19automod}. Researchers have iterated on these automated flagging tools, aiming to provide functionality that is better suited for context-specific communities; some examples include Crossmod \cite{chandrasekharan19crossmod}, which applies cross-community machine learning to identify problematic content \cite{chandrasekharan19crossmod}, and ModSandbox, which provides safer environments to test automated rules \cite{song2023modsandbox}.

To further complicate matters, Reddit has introduced new platform designs over time while continuing to support older ones, resulting in three primary interfaces available to users: Old Reddit, the current version of the Reddit.com site (referred to by long-time users as ``New Reddit''), and the Reddit mobile app. This is in part to cater to different user bases (e.g., longstanding users who prefer the older version versus relatively newer users who only know the current version). We discuss these interface differences with respect to the modqueue in \Cref{sec:modq}, as it provides further context for our study as well as informs survey design decisions. Beyond the native tools that Reddit provides, moderators frequently supplement them with third-party extensions such as Toolbox \cite{modtoolbox2024} or Reddit Enhancement Suite \cite{res2024}, as well as custom-built scripts that leverage the Reddit API \cite{reddit2017custombots,reddit2023moderationbots}.


Overall, prior research has established the importance of platform infrastructure and tooling in shaping moderation practices and challenges. Yet comparatively little work has examined the interfaces that support the critical task of mod review on Reddit, focusing instead on tools for flagging \cite{jhaver19automod,chandrasekharan19crossmod,song2023modsandbox} or rule development or co-creation \cite{zhang2020policykit}. Our study addresses this gap by focusing on Reddit’s modqueue, examining how moderators use it in practice and the challenges they encounter. In doing so, we aim to build a clearer understanding of how tools shape the work of mod review, and how focusing on this process can inform the design of moderation infrastructures that are more responsive to the diverse contexts and practices of online communities.

\section{Overview of Reddit's Moderation Queue}\label{sec:modq}

Reddit’s moderation queue, or \emph{modqueue}, is the platform’s central interface for consolidating items that have been flagged for moderator attention. These items include posts or comments reported by users, content filtered according to community-defined AutoModerator (Automod) rules (or other custom flagging tools), and material flagged by Reddit’s platform-wide automated spam-detection systems \cite{modqueue}. We use the terms \emph{items} and \emph{reports} interchangeably to refer to all flagged content, whether it was user-reported or automatically filtered.

From the modqueue, moderators can review reports and take actions such as approving or removing content, marking items as spam, locking comment threads, or applying flairs. While these actions can also be taken directly from subreddit pages, the modqueue is the only interface designed specifically to aggregate flagged items across all of a moderator’s subreddits, making it a critical locus for report review.

In this section, we provide an overview of the modqueue as it appears across Reddit’s three main interface versions, and describe how moderators often extend these official tools with third-party extensions. Our goal is to give readers the broader context necessary to understand our study design and, later, to interpret our findings.

\subsection{Modqueue Across Reddit Interfaces}

Reddit currently supports three main versions of the modqueue interface, corresponding to the three primary versions of Reddit itself, as outlined in the platform’s official documentation \cite{modqueue}. These include: (1) the Old Reddit interface, which retains the original modqueue design and has remained live but unchanged since Reddit’s broader site redesign began in 2018 \cite{pardes2018inside}; (2) the default version that is shown on reddit.com, which we refer to as ``New Reddit'' for the remainder of this paper, which features expanded moderation tools and more detailed content views; and (3) the mobile app interfaces for iOS and Android, which offer a streamlined version of the New Reddit modqueue with limited filtering and bulk action functionality. To contextualize our study, we briefly describe how the modqueue appears in each of these platform versions, and emphasize some notable differences in presentation and functionality. For more exhaustive documentation of available features, we refer readers to Reddit’s official support pages \cite{modqueue}.

\subsubsection{Modqueue on Old Reddit.} Old Reddit reflects Reddit’s pre-2018 site design \cite{pardes2018inside}; while Reddit no longer updates Old Reddit, it remains accessible and is still preferred by many long-time users. The modqueue here appears as a scrollable list of reports, with each item displaying basic information about the flagged content, such as the author name, number of reports, and karma score (see \Cref{fig:oldmq}). Moderators can take a small set of core actions directly from this view (e.g., mark as spam, remove, approve, or ignore future reports).

In addition to the main moderation queue, Old Reddit also provides several related queues that function as more targeted subsets: the Reports queue (items flagged by users), the Spam queue (content removed automatically or manually), the Edited queue (recently edited posts or comments), and the Unmoderated queue (submissions not yet acted upon). Together, these queues offer moderators different entry points into the same overall task of handling content that may require review \cite{modqueue}.
\ifthenelse{\boolean{anonymized}}{
\begin{figure}[htbp]
  \centering
  \includegraphics[width=\linewidth]{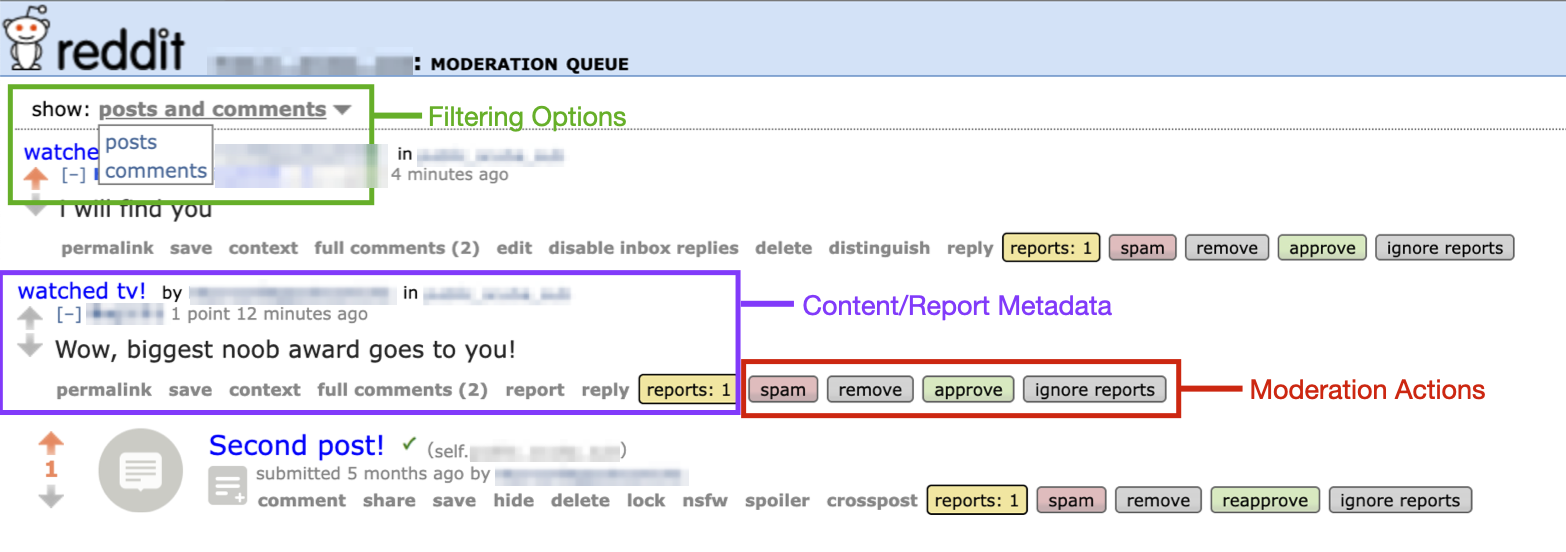}
  \caption{The Old Reddit modqueue interface. The modqueue appears as a list of items, which moderators can filter to show posts, comments, or both. Each entry displays basic metadata about the flagged content, including the content itself (e.g., comment text), the author, the number of reports, and minimal contextual information (such as the post a comment belongs to and the number of other comments on that post). Moderators can take core actions directly from this page, including approving, removing, marking as spam, or ignoring reports.}
  \label{fig:oldmq}
\end{figure}
}{
\begin{figure}[htbp]
  \centering
  \includegraphics[width=\linewidth]{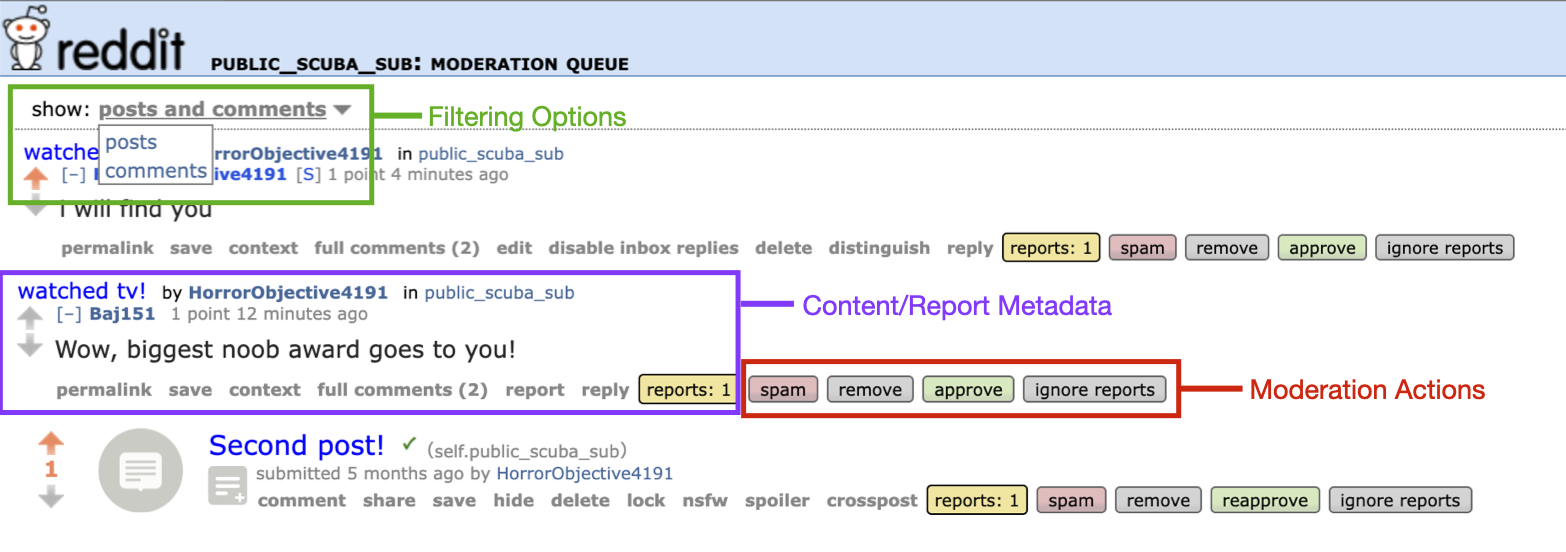}
  \caption{The Old Reddit modqueue interface. The modqueue appears as a list of items, which moderators can filter to show posts, comments, or both. Each entry displays basic metadata about the flagged content, including the content itself (e.g., comment text), the author, the number of reports, and minimal contextual information (such as the post a comment belongs to and the number of other comments on that post). Moderators can take core actions directly from this page, including approving, removing, marking as spam, or ignoring reports.}
  \label{fig:oldmq}
\end{figure}
}

\subsubsection{Modqueue on New Reddit.}The New Reddit version of the modqueue (\Cref{fig:modq,fig:newmq-interface}) retains the same underlying queue structure but expands the available features. The subqueues from Old Reddit are now presented as tabs within the moderation queue interface (\Cref{fig:newmq-interface}). Additional sorting and filtering options are available to allow mods to view more organized subsets of reports (see \Cref{fig:sortfilter}). 

Since Reddit’s 2018 site redesign, New Reddit has been the focus of ongoing platform updates, and many of the most significant changes to the modqueue have been introduced here (see \Cref{fig:newmq-interface}). Clicking on a reported item now opens a contextual panel alongside the queue, displaying report reasons and excerpts of surrounding discussion. A parallel feature allows moderators to click on usernames to open a user panel, which surfaces subreddit-specific information about that user (e.g., prior activity, local karma) and enables user-level actions such as sending a modmail, issuing a ban, or logging a note for other moderators. Together, these features aim to support more detailed review without leaving the modqueue, streamlining report triage and reducing the cognitive overhead of switching between tools \cite{reddit2024newmodqueueenhancements}. 

Another recent addition is the inclusion of real-time activity indicators \cite{reddit2024newmodqueueenhancements}, which signal when other moderators in the subreddit are acting within the modqueue. Reddit describes these indicators as intended to improve awareness and reduce the likelihood of conflicting or redundant decisions. It is unclear whether these indicators adequately present the actions of mods that may be utilizing the Old Reddit version of the modqueue.
\ifthenelse{\boolean{anonymized}}{
\begin{figure}[htbp]
  \centering
  \includegraphics[width=0.8\linewidth]{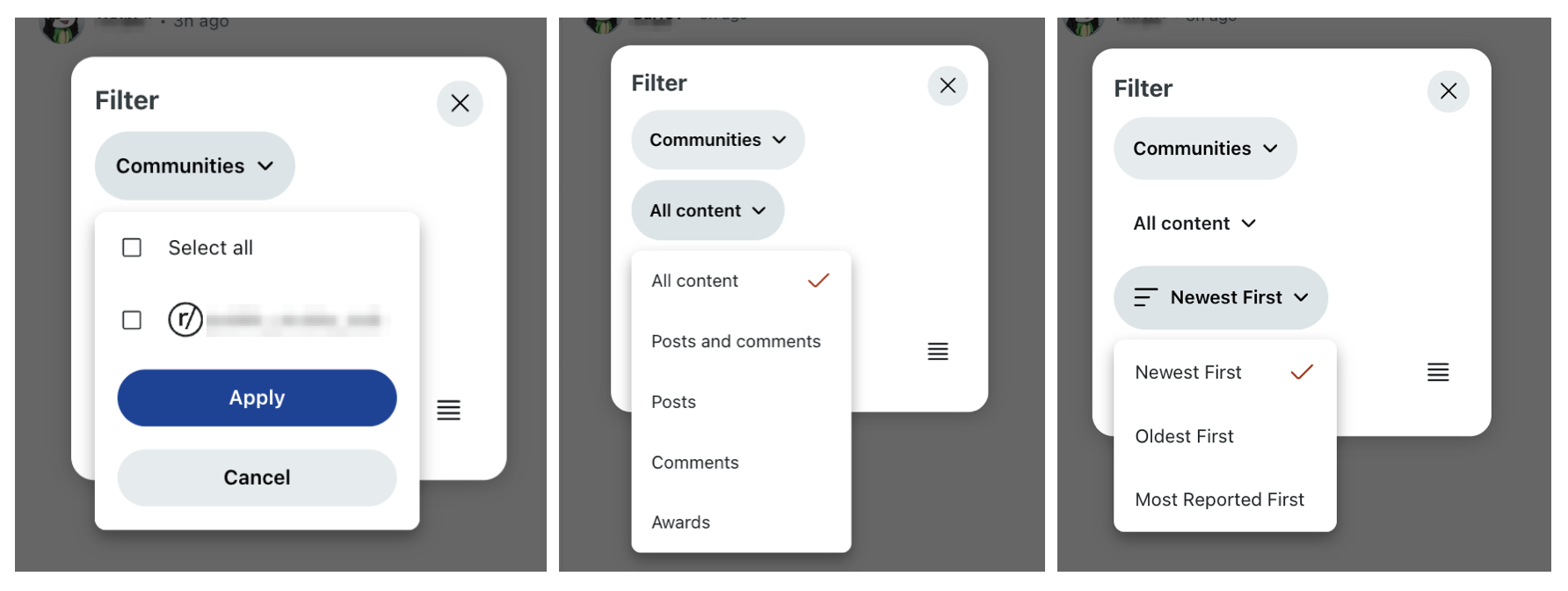}
  \caption{Additional filtering and sorting options in the new modqueue allows mods to view and prioritize certain report types over others.}
  \label{fig:sortfilter}
\end{figure}}
{
\begin{figure}[htbp]
  \centering
  \includegraphics[width=0.8\linewidth]{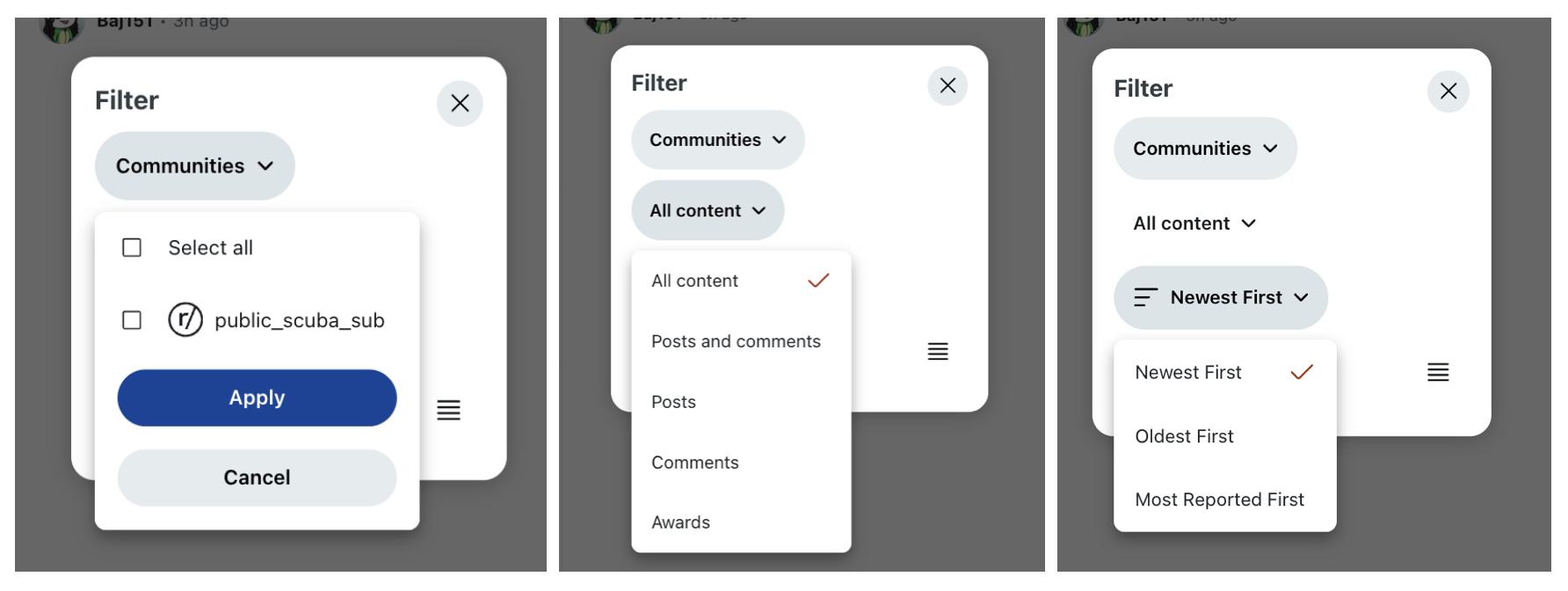}
  \caption{Additional filtering and sorting options in the new modqueue allows mods to view and prioritize certain report types over others.}
  \label{fig:sortfilter}
\end{figure}
}

\ifthenelse{\boolean{anonymized}}{
\begin{figure}[htbp]
  \centering
  \includegraphics[width=\textwidth]{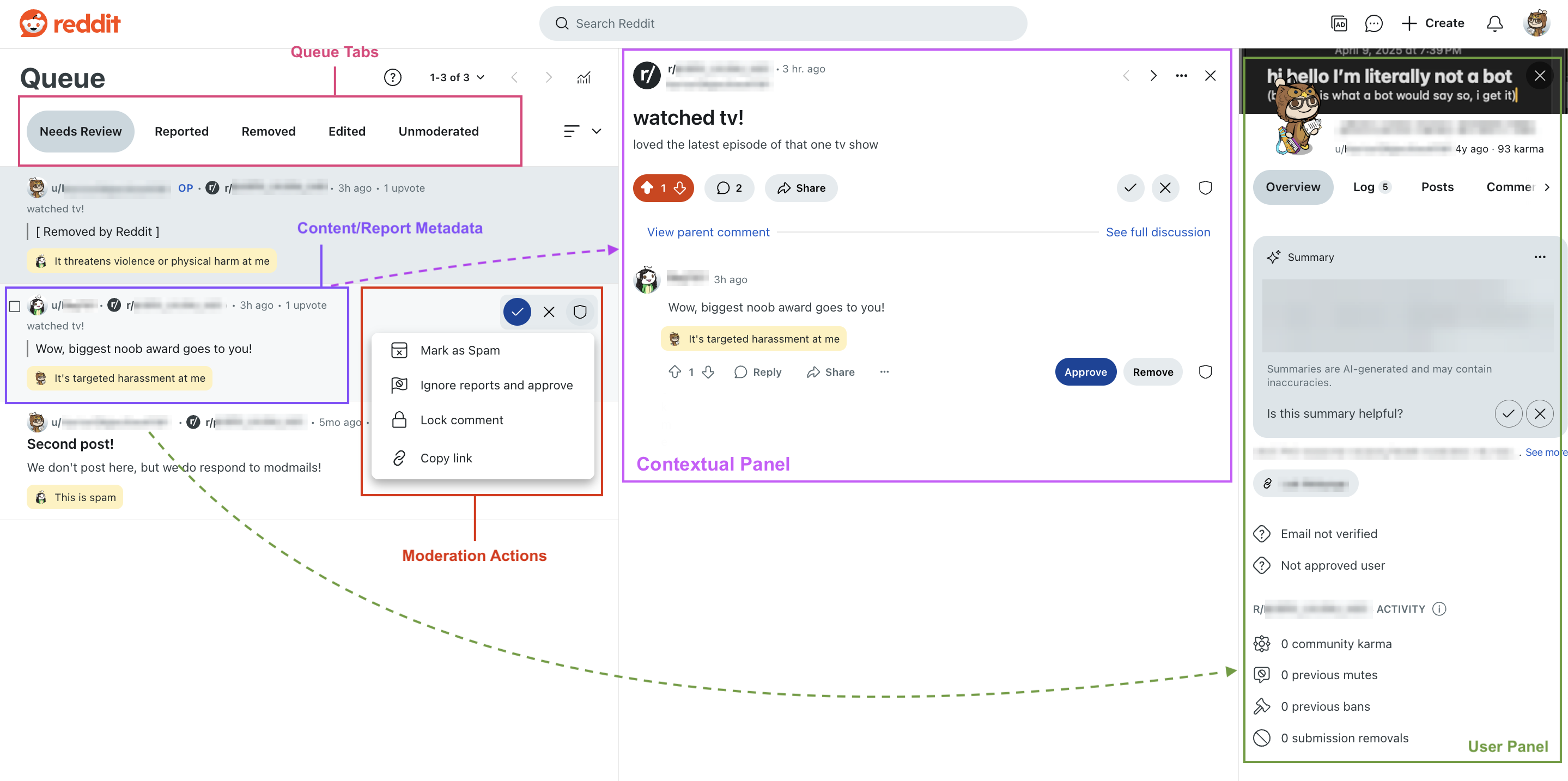}
  \caption{The New Reddit modqueue interface. Old sub-queues (e.g., Reported, Spam, etc.) are folded into the moderation queue as tabs. When an item in the queue is clicked, a contextual panel opens to the right that displays the surrounding discussion. When a username is clicked, a user panel opens to the far right that shows information about the user and their history in the subreddit.}
  \label{fig:newmq-interface}
\end{figure}
}{
\begin{figure}[htbp]
  \centering
  \includegraphics[width=\textwidth]{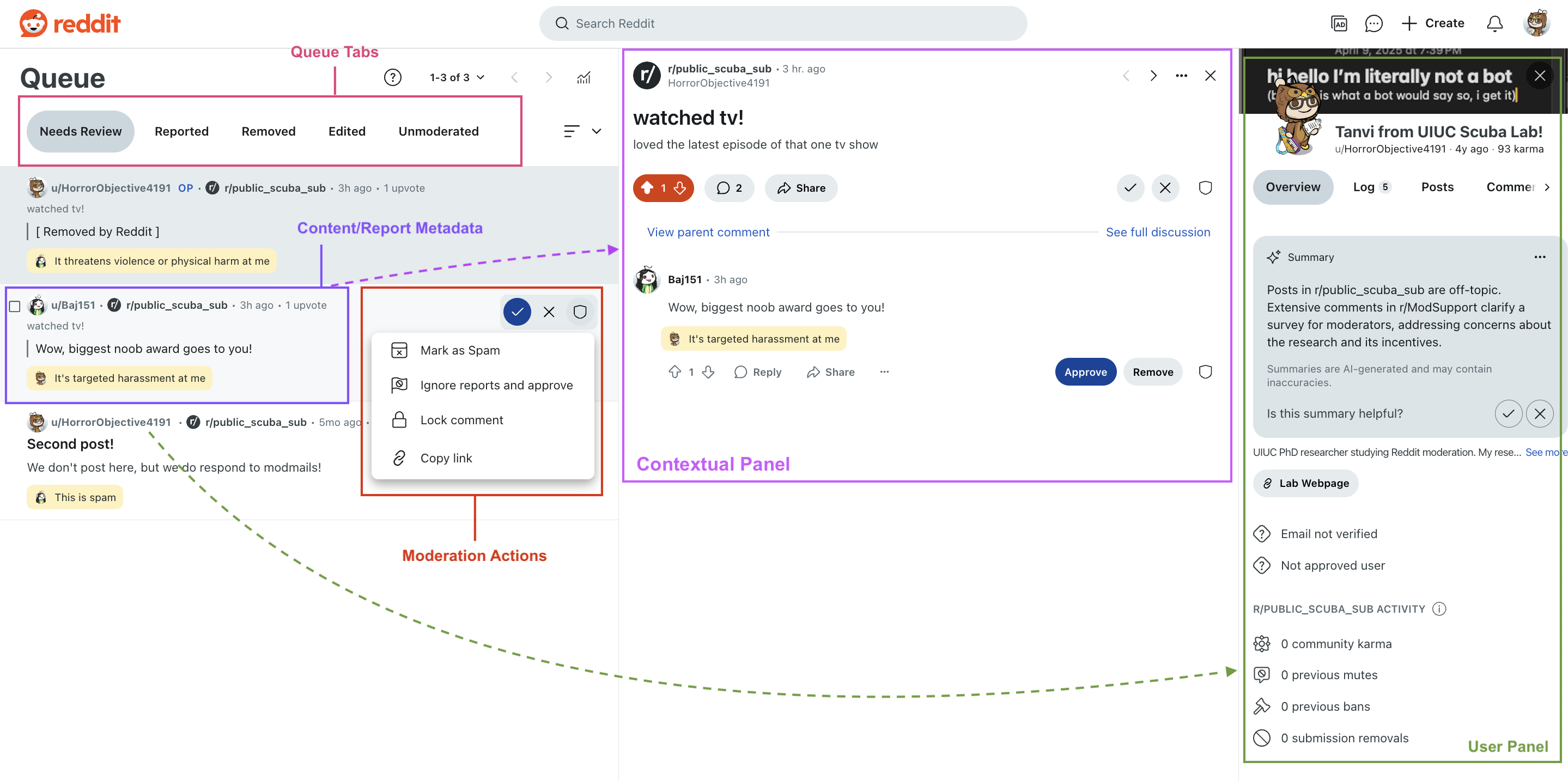}
  \caption{The New Reddit modqueue interface. Old sub-queues (e.g., Reported, Spam, etc.) are folded into the moderation queue as tabs. When an item in the queue is clicked, a contextual panel opens to the right that displays the surrounding discussion. When a username is clicked, a user panel opens to the far right that shows information about the user and their history in the subreddit.}
  \label{fig:newmq-interface}
\end{figure}}

\subsubsection{Modqueue on Mobile app.} Reddit’s mobile app offers an abridged version of the modqueue designed for “on-the-go” moderation. Reports appear in card form, and moderators can approve or remove items with swipe gestures, or long-press to access extended actions (e.g., lock comments, flair posts, mark NSFW). Bulk actions are supported in limited form, but advanced features such as contextual or user panels are absent.

\ifthenelse{\boolean{anonymized}}{
\begin{figure}[htbp]
  \centering
  \includegraphics[width=0.5\linewidth]{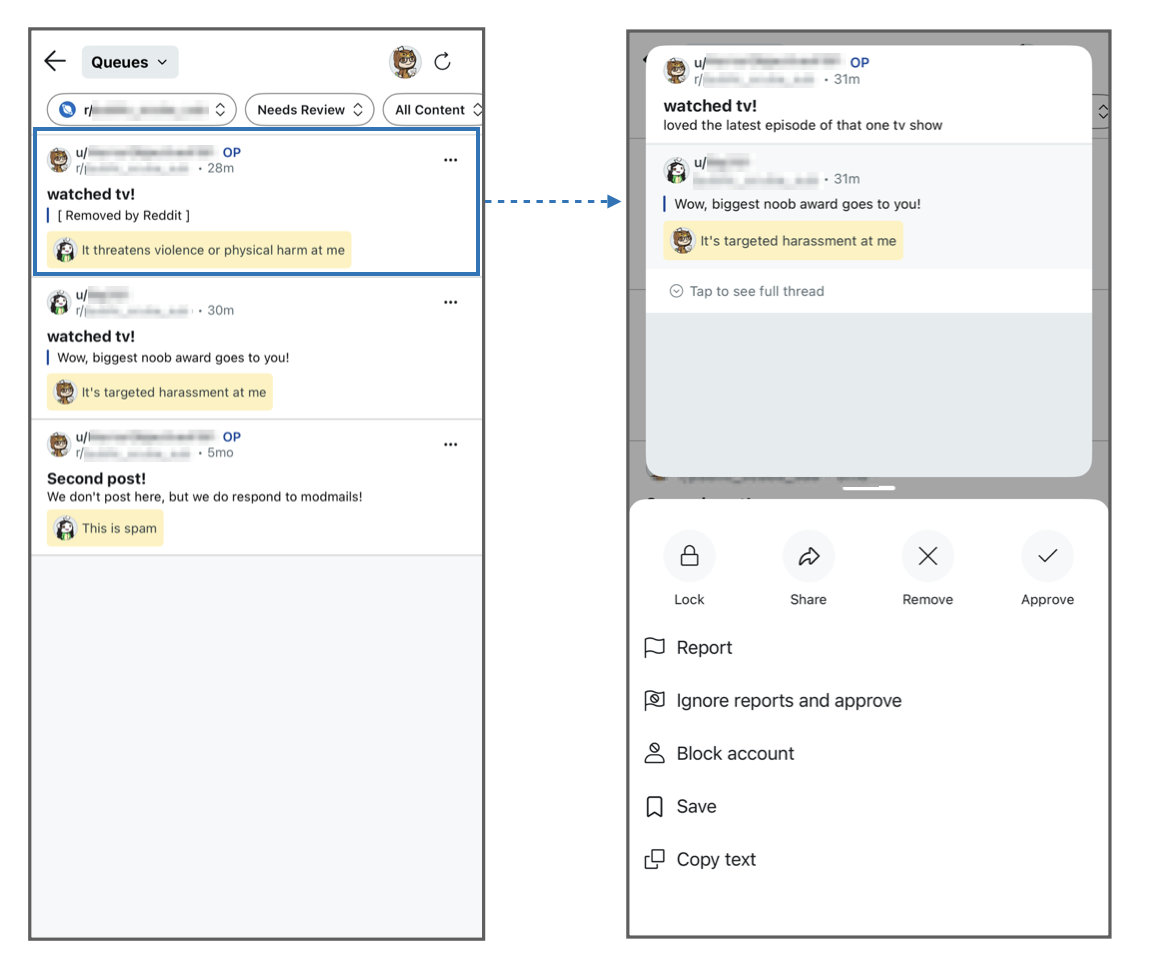}
  \caption{The modqueue on Reddit’s mobile app provides a streamlined version of the New Reddit interface. Reports are shown in a simplified list, and moderators can access moderation actions or additional information about each item by tapping on it.}
  \label{fig:mobile}
\end{figure}
}{
\begin{figure}[htbp]
  \centering
  \includegraphics[width=0.5\linewidth]{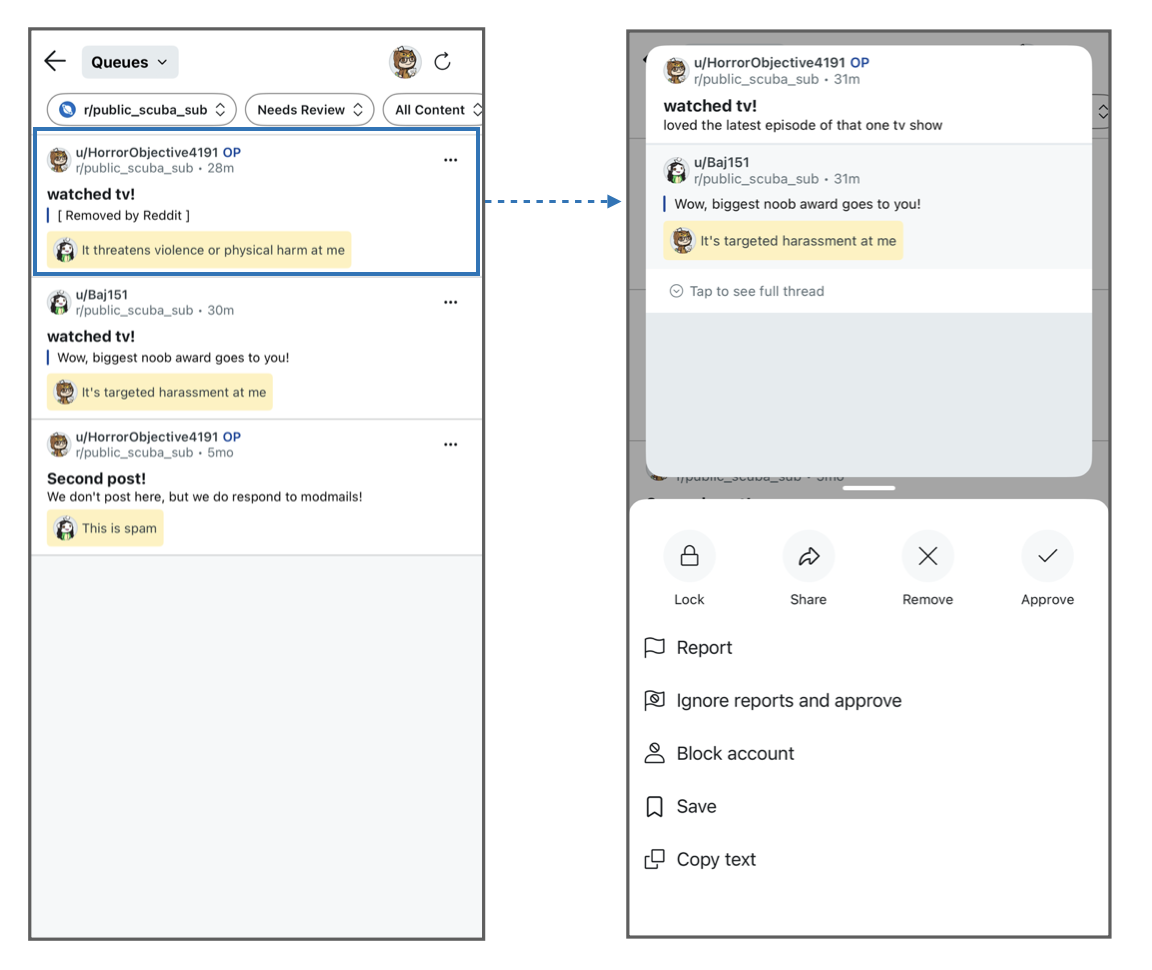}
  \caption{The modqueue on Reddit’s mobile app provides a streamlined version of the New Reddit interface. Reports are shown in a simplified list, and moderators can access moderation actions or additional information about each item by tapping on it.}
  \label{fig:mobile}
\end{figure}
}
    
\subsection{Third-Party Tools and Extensions}

As discussed in \Cref{sec:relwork:tools}, beyond Reddit’s official interfaces, moderators often rely on third-party tools to supplement or extend the modqueue. Popular examples include Reddit Enhancement Suite (RES), which offers customizable browsing and moderation shortcuts, and Mod Toolbox \cite{modtoolbox2024}, which provides additional moderation features that are unavailable in certain Reddit interfaces (or were formerly unavailable but later integrated by Reddit itself in newer versions) such as bulk actions, detailed user histories, and mod notes. These tools can fill gaps left by native interfaces, or restoring functionality that may be unavailable in Old Reddit or less accessible in New Reddit.

Because third-party tools and browser extensions might emphasizes different capabilities, moderators frequently piece together unique combinations of interfaces and extensions to suit their own workflows. For some, this means layering Toolbox features on top of Old Reddit; for others, it may involve using RES or running custom scripts built with the Reddit API to create alerts or notifications that show up as modqueue items that mods might view on-the-go using the mobile app. These practices reflect not only potential feature gaps in the official modqueue(s) but also the flexibility moderators exercise in tailoring their toolkits to suit their (and their community's) moderation needs.

\subsection{Summary}

In summary, across Old Reddit, New Reddit, and the mobile app, the modqueue consistently aggregates flagged items and provides moderators with a shared set of core actions (e.g., approve, remove, mark as spam, ignore reports). What differs across versions is how reports are presented and which contextual signals are surfaced. The Old Reddit modqueue offers a minimal, fixed list-based view with limited report information; the New Reddit modqueue provides several options for mods to re-sort or filter the list or reports, and introduces contextual panels, access to user-level actions, and real-time activity indicators; and the mobile app modqueue prioritizes speed and simplicity but omits many advanced features. In practice, moderators may also supplement these official interfaces with third-party tools, meaning that no single ``canonical'' modqueue experience exists. Instead, moderators navigate between versions and tools depending on device, habit, and community context. These differences provide essential background for our study, highlighting why moderator practices must be examined through moderators’ own accounts rather than inferred from the interface design alone.
\section{Methods}\label{sec:methods}

We conducted a survey of active Reddit moderators to examine how they engage with the modqueue in practice. In this section, we describe the survey design, recruitment strategy, and analysis approach.

\subsection{Survey Design}

Our survey was designed to capture the diversity of how moderators use (or why they choose not to use) with the modqueue. To accommodate the variation in moderation practices and tooling, we used a mix of multiple-choice, Likert-scale, and open-ended questions. Open-ended prompts were integral to our survey design, as they gave moderators room to describe modqueue practices and challenges in their own words, rather than being constrained by a predetermined framing of interface dynamics or workflows. Our goal was to capture richer, practice-grounded accounts from moderators, akin to the depth of interviews, while also reaching a broad and varied pool of participants. We acknowledge that this approach comes with certain limitations, which we discuss further in \Cref{sec:disc:lim}. 

The survey began with background questions about respondents’ moderation experience, the subreddits they moderate, the platform versions and tools they use, and how much of their time involves the modqueue. These questions provided context for interpreting variation in later responses. The remainder of the survey asked a variety of questions that could be categorized into four broad categories, each corresponding to one of our research questions, which we discuss below. Pilot testing suggested that our survey would anywhere from 15 to 25 minutes to complete, depending on what branches of the survey were accessed, and the level of detail provided in open-ended responses.

\subsubsection{Workflows (RQ1)} Because the modqueue may not be universally adopted or relied on in the same way across moderators, we sought to understand how it fits within their broader practices. We asked how frequently they use the modqueue, what kinds of tasks they carry out through it, and whether they combine it with other tools or workflows. For moderators who reported not using the modqueue, we asked how they handle reports instead, their reasons for avoiding the queue, and what changes might make them more likely to adopt it.

\subsubsection{Prioritization (RQ2)}
At its core, the modqueue is a list of reports presented in a predetermined order, but moderators can bypass that ordering and decide for themselves which items to review first. Interface differences add to this potential usage complexity: Old Reddit offers minimal sorting or filtering, while New Reddit introduces controls based on metadata such as report age or number of reports. This evolution in re-ordering functionality suggests that moderators may not simply process the queue as presented, but instead engage in \emph{prioritization}, selecting certain reports to handle before others based on their own judgments or report-specific information. To examine whether and how prioritization occurs in practice, we asked moderators to describe the strategies they use when deciding which item to review first. We also asked what advice they would give to a new moderator approaching the queue, as a way to surface potential variations in strategies that mod teams may view as useful or pertinent. Finally, we asked about the specific signals they pay attention to when selecting reports to review from the queue (e.g., item age, report source or volume, content type).

\subsubsection{Leaving the Modqueue (RQ3)}
Even though newer versions of the modqueue include features like contextual panels and user panes that are meant to surface the information needed for review, it is not clear whether moderators remain within the modqueue to complete review-related work. In other words, we want to guage whether mod review something that can (or should) be carried out entirely within the modqueue, or does it regularly require shifting to other parts of Reddit’s moderation infrastructure? To investigate this, we asked how often moderators leave the modqueue while reviewing reports and what exactly prompts them to do so.

\subsubsection{Challenges (RQ4)}

To surface the challenges moderators face when using the modqueue, we asked respondents to describe any frustrations they had experienced. This open-ended question was placed at the end of the survey, after respondents had already reflected on their practices through earlier questions, so that their answers would build on those reflections. In addition to this broad prompt, earlier in the survey we asked about mods' experience with (or lack thereof) a potential challenge implied by a recent feature addition: real-time activity indicators. This was a notable design choice given that moderation work is typically asynchronous (moderators are not expected to ``clock in'' and work at the same time), making real-time signals that track and broadcast moderator activity to fellow moderators within the modqueue particularly unconventional compared to other modqueue features. However, Reddit introduced the feature specifically to help ``reduce redundancy'' in modqueue review \cite{reddit2024newmodqueueenhancements}; this suggests that instances where two or more moderators act on the same report without awareness of each other (an event we dub as a ``collision'') might be a recurring issue.

\subsection{Recruitment}

We recruited moderators from a wide range of subreddits to capture diversity in experience, community type, and dependence on the modqueue. Because it was not possible to pre-determine which moderators relied most heavily on the modqueue, we aimed for breadth by contacting a large and varied pool, following strategies used in prior survey research on Reddit moderation communities \cite{lambert24positive}.

\subsubsection{Identifying Eligible Subreddits}

We began by identifying five public communities dedicated to moderation---r/ModSupport, r/modhelp, r/modnews, r/AskMods, and r/modclub---and collected usernames of Redditors who had posted in these forums between January 1 and April 1, 2025. For each user, we retrieved the list of subreddits they actively moderated. Subreddits were included if they met the following eligibility criteria, which was partly adapted from prior survey work conducted on Reddit: were at least three months old, had at least two human moderators, had at least 500 members, and had received average of ten or more comments per post in the last month.

We additionally chose to focus on English-speaking communities, and to do so used the \texttt{langdetect} tool \cite{langdetect} to analyze subreddit titles and descriptions. When language could not be reliably determined via \texttt{langdetect}, the first author manually reviewed subreddit content to make a final determination.

\subsubsection{Survey Distribution and Interruptions}

After filtering, we were left with a pool of 1,646 subreddits. We then began contacting the moderators of these subreddits via Reddit’s modmail system, which allows users to send messages to a subreddit's moderation team. Messages were sent in a rolling fashion over three weeks. 

Initially, we used a newly created Reddit account to contact subreddits, which was quickly shadowbanned by Reddit's automated spam-detection filter due to the volume of outgoing mail. To prevent further disruptions the distribution process, we resumed recruitment using a long-standing personal Reddit account (over two years old). To maintain transparency, we posted a public explanation of the account switch and responded directly to any moderators who raised questions about the legitimacy of the study. We also honored requests to exclude subreddits from our initial pool; this typically occurred when mods of a subreddit that had already been contacted replied to our modmail and requested that neither they or their ``sister'' (i.e. related) subreddits be contacted again. We additionally ensured that no subreddit received more than one message requesting participation. 

To encourage participation, we offered an optional raffle: one \$25 Amazon gift card was awarded for every 50 participants who opted in. This raffle-based scheme is commonly used in survey research with Reddit moderators \cite{koshy2023alignment,lambert24positive}.

By April 26, 2025, invitations had been sent to moderators of 1,400 of the 1,646 subreddits in our eligible pool. Recruitment was halted at that point due to an unrelated controversy involving researchers who conducted a study on Reddit without obtaining community consent, which sparked widespread concerns and distrust toward external research \cite{reddit2025cmvexperiment}. Out of respect for these concerns, we ended recruitment early.

\subsection{Analysis}

Our analysis aimed to surface the diversity of practices, strategies, and challenges moderators described, rather than to establish precise distributions or frequencies. Open-ended responses were qualitatively coded using an inductive, two-stage process: the first author conducted an initial pass to develop a provisional codebook, followed by a second pass in which the codes were refined and applied consistently. This process enabled us to highlight both common patterns and notable outliers, capturing variation across moderators without assuming any single dominant workflow.

For multiple-choice and numerical questions, we used basic descriptive statistics (e.g., counts and distributions) to provide background and context for the open-ended responses. We also drew on these responses to further situate notable accounts. In presenting the survey findings, our emphasis is on describing practices and patterns, rather than focusing on the exact prevalence of specific kinds of open-ended responses or themes.

\section{Findings}\label{sec:findings}

We first describe the backgrounds of our survey participants and the tools and interfaces they use (\Cref{sec:findings:backg}). We then turn to each research question in sequence: how the modqueue fits into moderators’ workflows (RQ1; \Cref{sec:findings:workflow}), how items are prioritized (RQ2; \Cref{sec:findings:priority}), why moderators leave the queue (RQ3; \Cref{sec:findings:leave}), and the challenges they encounter (RQ4; \Cref{sec:findings:challenges}).

\subsection{Respondent Backgrounds and Interface Use}\label{sec:findings:backg}

\begin{figure}[t]
    \centering
    \includegraphics[width=0.8\linewidth]{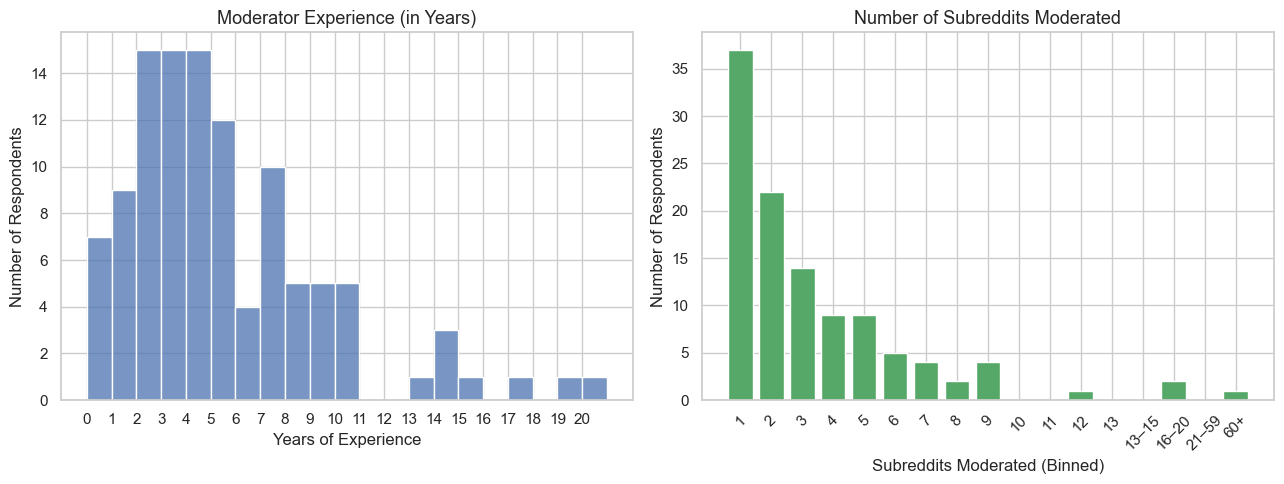}
    \caption{(left) Distribution of mod experience in years. (right) Distribution of the number of subreddits moderated by each respondent.}
    \label{fig:response_histograms}
\end{figure}

Our final respondent pool consists of 110 completed survey responses from moderators representing 100 distinct subreddits. Many of these moderators self-reported overseeing multiple communities: in total they reported actively moderating 408 unique subreddits, with a median of 2 subreddits per moderator, and one moderator reporting activity across 64 subreddits. \Cref{fig:response_histograms} (right) shows the distribution of subreddit counts.

Our survey respondents represented a wide range of moderation experience. Respondents had spent anywhere from less than one year to Reddit's full 20-year history as a mod, with the average respondent having 5.1 years of moderation experience. (\Cref{fig:response_histograms}, left) shows a histogram of the mod experience levels in years of our participants. Time spent moderating also varied: 43 respondents reported spending 1–5 hours a week on moderation work, 23 reported 6–10 hours, 20 reported 11–20 hours, and 14 reported more than 20 hours.

 Of the 408 subreddits mentioned by respondents, we were able to retrieve public-facing metadata for 386 communities; the remainder were private or otherwise inaccessible. 
 We found that the scale of these subreddits and the sizes of their moderation teams had substantial variation as well. With respect to the size of the subreddits: 42\% had over 100,000 subscribers, while 30\% had under 10,000. The majority of subreddits (59\%) have between 2 and 10 moderators, and a smaller subset (3\%) have teams larger than 30. We note that moderator team size reflects the number of listed moderators and may not account for inactive team members.

\subsubsection{Interface Preferences and Tool Usage}

Respondents reported using all three versions of the Reddit modqueue: 72 used the mobile app, 66 New Reddit, and 53 Old Reddit, with 67 reporting multiple interfaces. The most common combinations were New Reddit with mobile (34 respondents) and all three interfaces (17). Old Reddit users averaged 7.1 years of moderation experience, compared to 4.7 for New Reddit and 4.3 for mobile, suggesting greater resistence to New Reddit usage among veteran moderators.

Beyond native tools, 28 respondents reported using third-party extensions such as Mod Toolbox \cite{modtoolbox2024}, Reddit Enhancement Suite (RES) \cite{res2024}, Reveddit \cite{reveddit2024}, or develop custom bots using the PRAW API \cite{praw2024}. These tools offered bulk actions, additional filters, or user-history access, often filling gaps in Old Reddit but sometimes overlapping with features added to New Reddit. For example, P016 explained: \emph{``With modtoolbox, I'm choosing the option to expand all reports when browsing a specific sub's submission queue in Old Reddit. On New Reddit it's visible by default as well for me.''}

AutoModerator (Automod) was also frequently mentioned, not as a means for reviewing or resolving reports directly but as a source of items in the modqueue. Respondents described how Automod’s filters shaped the mix of items they encountered and sometimes influenced how they prioritized reports; we discuss this in more detail in \Cref{sec:findings:priority:source}. 

\begin{figure}[t]
    \centering
    \includegraphics[width=0.8\linewidth]{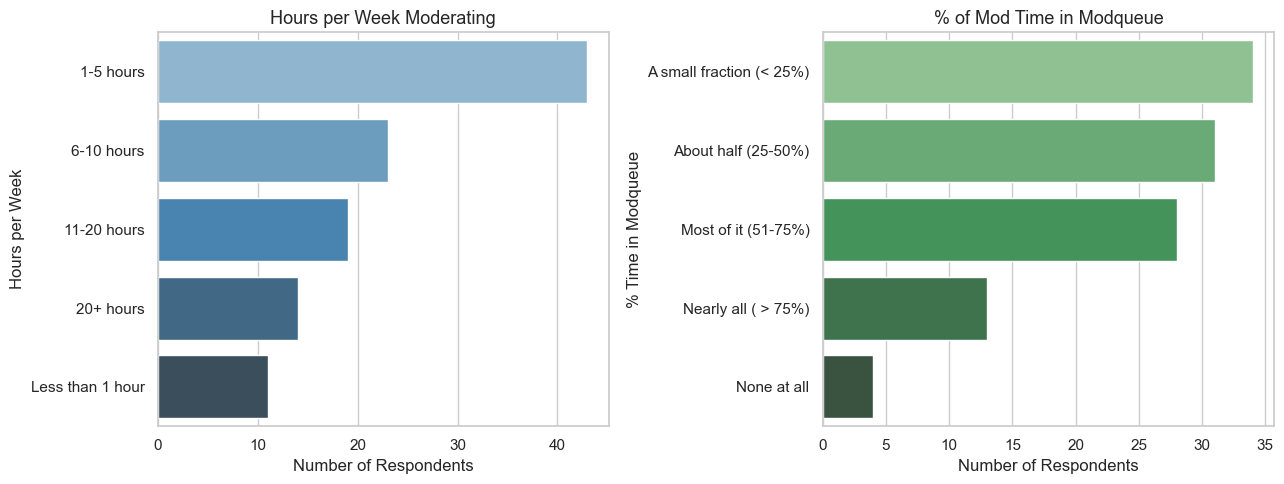}
    \caption{(left) Distribution of hours per week respondents spend moderating. (right) Proportion of time mods report using the modqueue.}
    \label{fig:modqueue_usage}
\end{figure}

\subsection{RQ1: How Does the Modqueue Fit into Moderators’ Moderation Workflows?}
\label{sec:findings:workflow}

The modqueue is designed as Reddit’s central workspace for reviewing reported and filtered items, but how it fits into moderators’ actual practices is less straightforward. Understanding whether, when, and how moderators rely on the modqueue provides a foundation for interpreting their later strategies for prioritization, reasons for exiting the interface, and challenges.

\subsubsection*{Summary} Overall, we found that respondents fit the modqueue into their workflows in pretty different ways. A small number avoided it almost entirely, relying instead on automation or lighter-weight in-thread actions (\Cref{sec:findings:workflow:no}). Most used it as a kind of checklist, working through filtered and reported items as part of their everyday routine (\Cref{sec:findings:workflow:every}). And a smaller set described using it more strategically, scanning for patterns or clusters of issues that signaled broader problems in the community (\Cref{sec:findings:workflow:less}). In practice, this means the modqueue is not a single, uniform workspace, but one that takes on different roles depending on moderators’ philosophies, the kinds of communities they moderate, and the types and volumes of issues those communities face.

\subsubsection{Skipping the Modqueue in Favor of Alternative Workflows or Automation}\label{sec:findings:workflow:no}

Only four of our respondents said they do not use the modqueue, but their reasons for bypassing it reflect broader concerns that may resonate with other Reddit mods who moderate similar communities, or who share similar backgrounds or moderation philosophies. These reasons included finding the modqueue inefficient or difficult to use, lacking the information needed to act confidently within the modqueue, feeling that their subreddits received too few reports to justify relying on it, or preferring a more lightweight, informal approach to moderation that didn’t involve switching to a separate interface.

One moderator, P078, who moderates two large subreddits, each with millions of members,  framed their avoidance of the modqueue in terms of how they viewed the role of moderation itself: \emph{``Moderator is a voluntary role. Having a designated place to do moderation feels odd. I like to think of myself as a user with more permissions, so I do not want a special page.''} This perspective further aligned with how P078 described setting up their communities: both subreddits rely heavily on Automod and custom AI-driven systems to handle most of moderation work. Rather than treat report review as a hands-on process, they saw automation as a way to reduce effort and let moderators focus on other human-centric moderation tasks: \emph{``Honestly, automate moderation away, moderators should work on community building.''}

\subsubsection{Everyday Uses: Reviewing Filtered Content, Responding to User Reports, and Clearing the Queue}\label{sec:findings:workflow:every}

Unsurprisingly, the most prominent use of the modqueue among the 106 moderators who reported using it was to review content automatically removed by platform-level filters or Automod. Several mods emphasized that this filtered content is not visible elsewhere, making the modqueue the only place where they can review and potentially release it to ensure that legitimate contributions are not mistakenly withheld from the community. Also expected, mods reported addressing user-submitted reports from the modqueue. Some moderators noted that this was not their primary reason for using the interface, and described responding to user reports more opportunistically during regular subreddit perusal, or only after reviewing filtered content first. 

Among moderators who used the modqueue primarily to review filtered or reported content, several described treating it as a checklist of action items, i.e., something they routinely cleared, or at least made steady progress on, as part of their day-to-day work. In communities with manageable report volumes or reports that were relatively straightforward to evaluate (often in subreddits with focused or lighthearted topics), this often meant working through the queue uninterrupted until it was fully cleared. Respondents also noted that this process was faster and more streamlined in the modqueue than by browsing the subreddit directly:

\begin{quote}
    \emph{I strive to keep a clear modqueue. Usually the modqueue is under 150 items so it is trivial to go through in under ten minutes. (P014)}
\end{quote}

\subsubsection{Less Obvious Uses: The Modqueue as an Activity Radar or Moderation Hub}\label{sec:findings:workflow:less}

Beyond the more expected uses of the modqueue, a few mods described their approach in ways that suggest they use it as a kind of activity radar, scanning for patterns, emerging issues, or signs of disruption to get a sense of the community’s temperature or surface potential problems. After spotting notable trends in the modqueue (e.g., clusters of similar reports or contentious comment threads that seemed to need broader investigation or intervention), these mods reported leaving the modqueue interface to take more targeted or multi-step actions directly from the content on the subreddit page. As one respondent put it (when offering advice on how new mods should approach the modqueue):

\begin{quote}
\emph{Think about it like you're a firefighter. You've got some smoke alarms going off [in the modqueue], check for patterns and clumps of alarms, that probably means a bigger fire. Prioritize those so you can lock or remove threads that shouldn't stay up any longer and hand out bans as needed, then go back [to the modqueue] and double check the smaller issues. (P019)}
\end{quote}

A handful of respondents also described being directed to or away from the modqueue during report review based on signals from elsewhere in their moderation workflow. These signals included Reddit push notifications, Automod alerts, modmail messages from users, or communication with fellow mods (sometimes on external messaging platforms). These signals shaped not only whether and when moderators entered the modqueue, but also what they focused on once inside it; we expand on how these signals influence report prioritization in \Cref{sec:findings:priority}.

\subsection{RQ2: Do Mods Prioritize Items Within the Modqueue, and If So, How?}\label{sec:findings:priority}

Although the modqueue presents reports as a list, moderators are not limited to processing items in the order shown. Depending on the interface, they can sort or filter by criteria like age or number of reports, or they can simply choose for themselves which items to handle first. We were interested in whether moderators stick to a sequential approach, or if they actively prioritize certain items based on their own judgments or community needs. In this section, we describe the different ways moderators talked about ordering their work in the modqueue, ranging from simple traversal strategies to more adaptive and multi-step forms of prioritization.

\subsubsection*{Summary} Our respondents reported using a variety of different prioritization strategies while addressing items in the modqueue. Some worked straight through the queue in the order items were presented in, while others used sorting features or their own judgment and observations to give precedence to certain items over others (\Cref{sec:findings:priority:seq}). These judgments often hinged on whether an item was flagged by a user or by Automod, and whether it was a post or a comment (\Cref{sec:findings:priority:source}). A smaller set of moderators described more adaptive strategies, first clearing quick or obvious cases before moving on to items that were older, more severe, or more likely to escalate (\Cref{sec:findings:priority:adaptive}). Overall, these responses suggest that prioritization in the modqueue is not uniform, but instead reflects moderators’ goals and values.

\subsubsection{Sequential or Priority-less Approaches}\label{sec:findings:priority:seq}

A little over half (56) of our respondents that reported using the modqueue described simple, sequential ways of moving through the modqueue as at least some part of the strategies they employ. Of these responses, many included mentions of traversing the modqueue in a particular direction, i.e., ``top to bottom'' or ``bottom to top'' (with no reference to how the modqueue may have been ordered. These responses often emphasized that what mattered most was addressing every item in the queue, regardless of order or prioritization. A handful of moderators echoed the view of P039, who explained that they had no specific strategy at all, seeing the goal simply as clearing the queue in its entirety: \emph{``All that is important to me is reviewing all reports. The order is irrelevant.''}

Some moderators who described sequential traversal of the modqueue specifically mentioned working from either ``newest first'' or ``oldest first.'' Those who preferred newest-first approaches often framed this choice as a way to prioritize activity that was happening in real-time, suggesting that more recent reports warranted quicker attention. By contrast, some of the mods that processed oldest items first, like P038, indicated that their ordering preference was motivated by broader-values like fairness: \emph{``I deal with the oldest queries first, as that seems the fairer option.''} Other reasons for preferring an oldest-first order, specifically for filtered items, was to avoid leaving potentially valid posts ``in limbo.'' 

\subsubsection{Prioritizing by Report Source or Content Type} \label{sec:findings:priority:source}

Another common way that moderators from our survey differentiated between reports was by whether they were submitted by users or automatically generated by filters or Automod. Several respondents said they checked filtered content first, often to quickly reinstate posts that may have been mistakenly removed. Other respondents felt strongly in the opposite direction and stressed that higher priority should be given to user-submitted reports, as they are more likely to reflect real concerns or active disruptions in the community. One respondent even offered contradictory guidance about whether to prioritize user-generated reports or content filtered by Automod, suggesting that newer mods should prioritize user-generated reports, but describing their own practice as starting with the filtered content. Both approaches reflect a need to prioritize the needs of users, but differ in whether the priority is placed on user-generated reports or user-generated content caught by automated filters.

\begin{quote}
\emph{Humans are always a priority over machines, so human-reported posts or comments should be tended to first. (P010, describing modqueue advice for new mods)} \\
\emph{I always begin with the ‘Requires Review’ queue, because these include content reported by our members and also AutoModerated content from new users who might have an urgent need. (P010, describing own modqueue workflow)}
\end{quote}

Prioritizing based on content type (i.e. posts vs. comments) was also commonly referenced. Some mods said they reviewed posts before comments, citing the fact that posts have higher visibility on the subreddit. Others prioritized comments, pointing to a greater likelihood of encountering personal attacks or escalating discussions that required intervention. These choices were often linked to perceived urgency based on risk of immediate harm or broader community impact.

\subsubsection{Multi-Step or Adaptive Prioritization}\label{sec:findings:priority:adaptive}

Rather than following a single rule or fixed order, a handful of moderators described using multi-stage or case-based prioritization strategies that adapted to the context or current needs of their communities. One type of strategy described in these types of responses was to clear the most obvious or easily resolvable reports first, then switching to more nuanced or potentially high-impact items: 

\begin{quote}
\emph{Clear the most obvious and uncomplicated reports first. Then switch to a FIFO method and try to address reports oldest to newest but with severity of issues taking priority. (P058, describing modqueue advice for new mods)}
\end{quote}

These multi-step strategies sometimes combined modqueue-enabled features with the respondents’ own mental categorizations of the queue. For example, P081 described scanning for obvious issues among posts first, then mentally sorting the remaining items based on factors like content status and how the item was flagged.

\begin{quote}
\emph{I first search Posts in the modqueue.  I scan through the list looking for obvious problems.  I mentally prioritize posts that have gone live and have comments.  Then I look at posts being held by Crowd Control\footnote{Crowd Control is a moderation feature on Reddit that automatically filters or restricts comments from users based on their account status (e.g., if a user is very new or has very low karma) designed to automatically filter or restrict comments on posts based on the commenter’s account.} or by the automod. (P081)}
\end{quote}

Some of these adaptive or multi-step prioritization strategies referenced a desire to prevent future moderation work. Reports on potentially contentious or high-visibility content were sometimes addressed early to avoid escalation, reduce repeated reports, or preempt user complaints.

\subsection{RQ3: Why Would Moderators Need to Leave the Modqueue While Reviewing Items?}
\label{sec:findings:leave}

The modqueue, and especially the New Reddit version, attempts to give moderators everything they need to review reports in one place. However, in reality, moderator review is rarely so contained: the meaning of a comment can depend on its surrounding thread, a post may only make sense in light of a user’s past behavior, and moderation decisions often draw on cues beyond what the queue itself displays. This raises the question of whether the modqueue can truly support end-to-end moderator review, or if moderators must regularly step outside the interface to do their work.

\subsubsection*{Summary} In our survey, over 84\% of respondents said they ``sometimes,'' ``often,'' or ``almost always'' leave the modqueue while reviewing a report to seek additional context. Our open-ended responses show that, despite newer features that provide contextual or user-specific features, many mods leave the modqueue to gather this information, or other external information, to help make moderation decisions. Moderators routinely step out to check the broader conversation (\Cref{sec:findings:leave:context}), investigate a user’s history (\Cref{sec:findings:leave:user}), or to consult previous moderation actions, either by reviewing the moderation log or by consulting fellow community moderators (\Cref{sec:findings:leave:mod}). In some cases they even leave Reddit altogether to fact-check or verify information (\Cref{sec:findings:leave:ext}). Notably, even moderators who used the New Reddit modqueue---which offers features like contextual panels and user history previews (see \Cref{sec:modq})---still described leaving the interface to gather the same kinds of information.

\subsubsection{Reviewing the Broader Conversation Context}\label{sec:findings:leave:context}

The most common type of context our respondents reported leaving the modqueue for was that of the surrounding conversation. This context was often necessary for interpreting tone, understanding how a specific comment fit into an ongoing discussion, or recognizing escalation patterns that weren’t obvious from the reported item alone.

Mods also described using this opportunity to identify other posts or comments in the same thread that needed attention. In some cases, leaving the modqueue to investigate a single report led to broader moderation work, echoing earlier descriptions of the modqueue as an activity radar (\Cref{sec:findings:workflow:less}):

\begin{quote}
\emph{In general, if I find a comment that needs to be removed for a rule violation, I'll click through to read the thread and get the full context. It won't change the need to remove the comment that appeared in the queue, but it's likely if one person is ignoring a rule I'll find more people doing the same within that conversation. (P086)}
\end{quote}

\subsubsection{Investigating User Behavior Patterns and Implementing User-Specific Actions}\label{sec:findings:leave:user}

Another common reason moderators left the modqueue was to investigate the history of the user who submitted the reported content. This step was important for assessing relevant patterns of behavior that might alter how content is interpreted and what user-specific actions should be taken. This included evaluating whether the reported content seemed like a one-off lapse (potentially warranting a warning), a repeated issue (justifying a suspension or stricter action), or part of a broader pattern of trolling, spam, or boundary-pushing behavior (which might lead to a permanent ban).

Mods differed in how much of a user’s behavior they felt was relevant. For instance P084, stated that \emph{``If [a user is] reported for a 'borerline' [sic] comment, maybe something that *Could* be sarcam [sic] or a joke, I like to look if they have a history of agitation or trolling, not just in my sub, but in others.''}. On the other hand, P072 stressed that they only review a user's activity within the subreddit, saying they \emph{``do not punish a person for behavior outside of the sub.''}

Beyond helping moderators interpret user intent or behavior, some user-specific actions such as banning still required leaving the modqueue entirely. Many respondents reported only leaving the modqueue page to go to the `ban users' page, suggesting that certain moderation actions still remain unsupported within even the New Reddit modqueue interface.

\subsubsection{Looking Up Moderation History and Seeking Input from Other Mods}\label{sec:findings:leave:mod}

In some cases, mods left the modqueue to look up prior moderation actions to either better understand how similar content or cases had been handled or to check what past actions had already been taken towards relevant reports or users. This often involved consulting the moderation log\footnote{The moderation log, or modlog, is a record of all moderation actions taken within a subreddit, including removals, bans, and approvals. It helps moderators audit actions taken by themselves or others \cite{reddit_modlog}.} (modlog) or checking user-specific mod notes (a feature enabled by Mod Toolbox \cite{modtoolbox2024}). Sometimes mods reached out to their moderation team for additional input. Respondent P024 described that their moderation team typically \emph{``[uses] Discord or mod notes...[to] often ask for a second opinion, or context from past mod actions.''} The practice of consulting other mods during report review from the modqueue was common but not universal: in response to a direct survey question about whether they consult fellow mods while reviewing, 58\% of moderators said they do, while 38\% said they do not.

In some cases, mods admitted to skipping certain reports as a way to leave them for other, perhaps more experienced, mods to handle; some respondents indicated that they leave notes or send messages to other mods to signal that they had skipped over the report, while others made no mention of taking any additional action besides skipping things that they were not confident about. For instance, P031 admitted \emph{``Sometimes I’ll leave something for another mod to weigh in on. Better to wait than to act too fast and regret it.''}

\subsubsection{Leaving Reddit for Verification or External Research} \label{sec:findings:leave:ext}

In a few cases, mods described leaving not only the modqueue but the Reddit platform itself to verify claims or better understand the reported content in question. This included using external tools or search engines to identify references, check facts, or interpret unfamiliar material in order to make appropriate moderation decisions. For example, P088, who moderates a subreddit pertaining to books, supplied \emph{``I may need to research whether or not a cover is for a real book, or if it’s an actual real cover. For that, I exit Reddit and use google and other search resources.''}

\subsection{RQ4: What Challenges Do Moderators Face When Using the Modqueue?}
\label{sec:findings:challenges}

If the modqueue is supposed to be the central place for moderators to review reports, then its utility also depends on how well it supports the realities of their work. Prior findings showed that moderators rely on the modqueue in different ways, and also hinted at moments of friction, i.e., times when the interface didn’t line up with what moderators needed or expected. Here we look more closely at the kinds of obstacles moderators described when working in the modqueue, from coordination issues to limits in the design itself.

\subsubsection*{Summary} Across our survey responses, moderators pointed to recurring pain points in how the modqueue functions or how it has evolved over time. Coordination issues that lead to collisions remained common, even to respondents that used New Reddit, which has real-time activity indicators (\Cref{sec:findings:challenges:col}). Other challenges faced in using the modqueue were difficulties in enacting multiple actions in report to an item (\Cref{sec:findings:challenges:dis}). Sorting, filtering, and logging features were sometimes described as buggy or inadequate, leaving moderators without adequate signals to support their review work (\Cref{sec:findings:challenges:dis}). Finally, the constant need to juggle multiple interface versions and third-party tools added frustration, especially when Reddit’s own updates broke long-standing workflows (\Cref{sec:findings:challenges:inter}).

\subsubsection{Coordination Issues and Collisions}\label{sec:findings:challenges:col}

Nearly three-quarters of respondents (74.5\%) reported experiencing a collision---i.e., instances where multiple mods are simultaneously working or respond to the same report---while working in the modqueue, while only 13.6\% said they had not; the remainder were unsure of whether or not a collision had occurred to them or not. The main reason mods' that reported never having experienced collisions had cited all had to do with things that limited the number of mods that were working on the modqueue at the same time (e.g. having low report volume and correspondingly smaller or less active moderation teams). One mod, P010, reported that \emph{``Mods can see whether there are other mods actively moderating the queue, at any given time. There is generally no need to have more than one mod at a time in the queue;''} their response indicates that they do make use of the real-time activity indicators as a way to decide whether they should conduct review via the modqueue. 

Most respondents that reported experiencing collisions described it as being a rare or occasional occurrence. Many of the mods that mentioned experiencing collisions frequently were responsible for moderating large subreddits with large moderation teams. Many of these respondents expressed frustration or annoyance surrounding the frequency at which they (and their moderation teams) were experiencing collisions. For instance, P065's stated that collisions \emph{``[happen] frequently. It’s annoying because we almost always have a backlog so if we’re both working on the same items, we’re wasting time.''}

On the other hand, P064, had a perspective on collisions that shed light on certain UI issues surrounding features that shed light on other moderator activity, as well as the common ways mods like themselves found out about collisions: 

\begin{quote}
\emph{Collisions are pretty frequent. It's hard to estimate exactly. There are UI indications about other moderator activity in the queue, but they are so subtle and unreliable that I don't pay attention to them. Usually I find out when I go to issue a ban and the user is already banned, or a double mod comment appears. (P064)} 
\end{quote}

Some mods said they take extra precautions (such as refreshing the queue more often, working in reverse order, or starting from the middle or opposite end of the modqueue) specifically to avoid such collisions. When collisions result in conflicting actions (e.g., one mod removes a post while another tries to approve it), teams often turn to Discord or modmail to coordinate after the fact, either to align on a final decision or to talk through different moderation instincts. These moments were sometimes framed as learning opportunities, especially for newer mods:

\begin{quote}
    \emph{[Collisions happen] every now and then. I recently recruited new trial mods for [subreddit], so it's always fun to see them working, and check on their progress. Sometimes we make different calls, and then we'll discuss it in the mod discord, and use it as a teaching opportunity, or I'll take feedback and change how I do things. It's not too often, though, because I worked hard to build a team who are awake in different time zones... (P019)}
\end{quote}

P019's response also touches on how mod teams that are naturally ``spread out'' with respect to moderator activity are able to naturally avoid or minimize the issue of collisions during modqueue use. 

\subsubsection{Modqueue Items Disappear Before Mods Are Fully Done With Them}\label{sec:findings:challenges:dis}

Another kind of disruption happens when mods need to take multiple actions on (or stemming from) one modqueue item, only for that item to disappear from the queue after the first action is submitted. This is especially frustrating in cases where a report requires more than one step (e.g., removing a post, locking the thread, and banning the user):

\begin{quote}
\emph{It's annoying that as soon as I take an action (even minor action) on an item in the queue that the item disappears... Sometimes I need to take more than one action, for example deleting the offending comment, AND locking the thread, AND banning the person. (P006)}
\end{quote}

Because items vanish from the modqueue once acted upon, completing these follow-up steps often requires navigating elsewhere, which interrupts mods' modqueue workflow. Participant P105 said \emph{``Sometimes I second guess myself and it's tricky to backtrack.''}

\subsubsection{Modqueue Signals and Features That Fall Short}\label{sec:findings:challenges:feats}

Certain modqueue sorting and filtering features were reported as being unreliable or not adequate enough to support the actual functionality respondents actually wanted. For example, a few mods said that sorting by the number of reports, a New Reddit modqueue feature \cite{modqueue}, didn't actually work well: 

\begin{quote}
\emph{At times in the past, I have tried to sort by posts that have multiple reports, but that doesn't work very well. I wish I could see human-generated reports first, with automod reports later, but that doesn't work. So I usually end up just using a sort order that puts the newest posts first. (P081)}
\end{quote}

Some mods also criticized the overall lack of transparency or accuracy in how how certain report categories determined or displayed. Spam reports in particular were flagged as unhelpfully collapsed under a single label, even though they can come from different user-selected reasons; as P025 stated \emph{``Spam reports show up as just ‘spam’... but they’re actually several different options... very unhelpful.''}

In other cases, mods described noisy or inconsistent mod logs that made it harder to trace what had happened to a piece of content. This was particularly frustrating for mods who were trying to identify repeat behavior or interpret past actions. P086 described the following issues with how Reddit keeps track of the modlog for individual users: 

\begin{quote}
\emph{There's a lot of mod log noise when viewing a user's mod log from the mod queue... There can be up to three separate mod log entries for the same content... Having triplicate entries in the mod log makes it a royal pain to read the user's history to find out if they're a repeat offender. (P086)}
\end{quote}

Finally, some mods raised concerns about being left without tools to handle malicious or bad-faith reporting. In particular, a few respondents called for more robust (or simply more) ways to detect and respond to coordinated or abusive use of the report feature, such as surfacing groupings of reports or providing meta-data about reporters to more quickly identify and act on abusive reporters. 

\subsubsection{Too Many Interfaces, Not Enough Tool Integration}\label{sec:findings:challenges:inter}

Several moderators expressed frustration with having to juggle multiple interfaces and third-party tools to carry out what they saw as basic moderation workflows. While many relied on browser extensions like Toolbox or RES to supplement missing features, they wished that more of these capabilities were built into Reddit directly, especially for mobile users or less tech-savvy moderation teammates. For instance, P057 stated \emph{``The toolbox makes things nice, would be nice to have that all built directly into the site and the app.''}

Some respondents pointed to specific moderation features that are available through third-party tools that have become harder to use as they are not supported in newer versions of the Reddit UI. As Reddit’s UI has evolved, respondents reported struggles to keep up with the vast amount of changes, especially when long-standing workflows break due to changes that disrupt the functionality of extensions or older features:

\begin{quote}
\emph{Reddit keeps changing the UI, and has forced an update to the newer UI where moderation extensions don't work, and so moderating now is more cumbersome than it was 2 years ago. (P061)}
\end{quote}

\section{Discussion}\label{mqs:sec:disc}

Our study surfaces the diverse, strategic, and sometimes improvised ways in which Reddit moderators use the modqueue to conduct mod review in practice, as well as the challenges they face while doing so. In this section, we discuss the main takeaways of our findings, discuss implications for feature and platform design, and highlight limitations and directions for future work. 

\subsection{Key Takeaways}

\subsubsection{Human Oversight of Automated Decisions Still Relies on the Modqueue}

While reports can be reviewed elsewhere, the modqueue remains critical for any subreddit that uses Automod or other filtering tools. Because filtered content is only accessible through the modqueue, it becomes the primary venue for surfacing potential false positives or reviewing borderline cases. In this sense, the modqueue is not just a triage interface, but a key point where human judgment gets reasserted in workflows that rely heavily on automation. Mods who trust automation fully (and prefer to review reports while browsing the subreddit) may bypass it, but those who aim to strike a balance between automated filtering and human decision-making will still depend on the modqueue.

\subsubsection{Interfaces Shape Review Workflows, But Mods' Values Shape How Interfaces Are Used}

Mods often rely on interface-specific features (such as certain orderings, filters, or tabs) to organize their review work on the modqueue. These choices usually reflect broader goals: filtering to posts might help surface more visible content, while filtering to comments could highlight more personal or aggressive behavior; sorting by newest- or most-reported-first might surface time-sensitive issues, while sorting by oldest-first can reflect a commitment to fairness. Even default sequential traversal often reflects a desire to ensure all items get addressed. These patterns suggest that interface affordances become proxies for broader community or moderation values. Understanding those values more explicitly could point to additional design directions.

\subsubsection{Mods Leave the Modqueue to Gather Context Despite Built-In Features}

Reviewing reports and making moderation decisions often requires information beyond what is shown in the report itself. Reddit’s introduction of contextual panels in the New Reddit modqueue suggests an awareness of this, yet many mods still reported leaving the modqueue to gather necessary context. This includes checking surrounding conversations, viewing user histories, or looking up past mod actions—all of which are technically supported by contextual panels. That mods still leave the interface suggests either a lack of awareness, a lack of trust, or that New Reddit is not their default interface for deep review.

\subsubsection{Fragmentation and Tool Incompatibility Can Disrupt Workflows}

Most mods in our survey reported using more than one version of the modqueue, often supplemented by third-party tools. This mix of interfaces can cause problems. For instance, the activity indicator in New Reddit's card view only works if everyone on the team is using New Reddit, which is not always the case. And while some mods prefer New Reddit, others stick with Old Reddit because it is more compatible with long-used extensions. As Reddit updates its UI, compatibility with third-party tools can break, creating tension between interface updates and workflow reliability. While offering multiple options gives mods flexibility, the lack of coordination or integration across these options can cause friction.

\subsection{Implications for Platform and Moderator Tool Design}

\subsubsection{Make Moderation Features Lightweight and Configurable}

Reddit has added features like contextual panels and activity indicators to help mods access necessary information for review and circumvent recurring issues surrounding coordination. But these features only matter if mods know about them and actually choose to use them. Some may find them unintuitive, unnecessary, or unavailable in their preferred interface. Instead of adding complexity, Reddit might consider offering lightweight or customizable versions of these features. For example, instead of a full contextual panel, mobile or Old Reddit users could benefit from simpler inline previews showing just parent comments or key metadata, adding to the existing content provided in the Old Reddit view of the modqueue, without majorly disrupting the minimalistic layout. Letting mods set preferences for how much context they see could make these features more compatible with existing workflows, and could potentially prompt more users to make use of them.

\subsubsection{Supporting Modular Tooling Across Interfaces and Subreddits}

Because mods use a mix of tools and interfaces, platforms like Reddit should prioritize compatibility and modularity. Allowing mods to mix and match features would make it easier to create workflows that work for different communities and individuals. This matters even more for mods working across multiple subreddits with different needs. 

Subreddit-specific interface settings could also help mods adapt to these differences, and ensure that coordination tools (like activity indicators) function as intended. For instance, if one large subreddit wants all mods using the activity indicator, they should be able to standardize that experience, specifically for review of that subreddit's content, to improve coordination among their mods. This may push back on mods existing, personal workflows, but a balance must be struck between accommodating the needs of individual mods and the needs of mod teams and communities. 

\subsubsection{Providing Functionality for Pattern-Recognition}

Some mods described using the modqueue not just as a place to process items, but as a way to scan for larger issues—like patterns in reports or spikes in problematic content. This repurposing reveals an underutilized strength of the modqueue: it centralizes signals from across the subreddit. When reports are timely and accurate, this can help mods spot disruptions that wouldn’t be obvious from isolated posts. Currently, this kind of pattern recognition is not directly supported. Mods rely on instinct and experience, rather than interface features, to detect trends. Some third-party tools support tagging or grouping, but, as previously mentioned, those tools can break when Reddit’s UI changes. This makes pattern-based use of the modqueue harder for less experienced mods to adopt.

Future design work could help by letting mods group, annotate, or track related reports more easily.
This could involve augmenting the modqueue with features tested in recent work on visualization tools to guide moderator attention and summarize trends/patterns in online activity~\cite{choi23convex, liu2025needling}.
More ambitious ideas include building interfaces or tabs that surface or visualize trends, or consolidate other moderation signals, such as post frequency, user behavior, or modmail alerts. 
These features could help mods respond proactively, rather than reactively. This kind of attention management may be even more impactful to broader moderation efficacy than just improving functionality for item-by-item review.

\subsection{Limitations and Future Work}\label{sec:disc:lim}

\subsubsection{Limited Depth and Lack of Follow-Up in Survey Responses}

Our survey reached a broad and diverse group of Reddit mods, but it did not allow for deeper follow-up or clarification. Because we prioritized breadth over depth, open-ended responses were often concise and lacked the detail that longer interviews or probing questions might have surfaced, and we could not resolve ambiguities in participants’ answers. Also, while we asked about the subreddits mods currently moderate, we did not ask them to tie specific practices to specific communities. As a result, we could not precisely relate modqueue usage to subreddit characteristics like size, topic, or structure, unless mods referenced these in their responses. Future work could address this limitation by conducting interviews with mods from subreddits of different types, to get a more targeted understanding of the factors that incite certain modqueue practices.

\subsubsection{Biases in Self-Reported Workflows}

Like any study based on self-reported data, our findings reflect how participants interpret and describe their own actions. Some may simplify workflows, omit edge cases, or describe what they think is best practice rather than what they actually do. Without direct observation or platform-side data, we cannot assess how well these descriptions match day-to-day mod activity.  

Future work could further contextualize or expand upon our findings using other, less biased methods. For instance, collecting lightweight forms of usage data (e.g., how mods sort the queue, how often they leave the interface, or which items they interact with) alongside data regarding community behavior and report activity could provide a clearer picture of how modqueue workflows unfold in practice. That said, many mods might find this kind of tracking invasive, and such concerns could limit the types of subreddits and mods who would be willing to participate. Still, even limited usage data could help identify points of friction or divergence between how the modqueue is intended to be used and how it is actually used.

\subsubsection{Lack of Interface-Specific Insight}

Although we intentionally kept our survey questions high-level to accommodate multiple interface configurations, this choice limited our ability to analyze how specific modqueue versions (e.g., Old Reddit, New Reddit, mobile app) shape moderator behavior. Many respondents used multiple interfaces in parallel, and some may not have known which features were specific to which interface. This ambiguity makes it difficult to tie reported practices or frustrations to particular interface designs or features (unless features mentioned were exclusive to one interface). As previously mentioned, future work could address these limitations by complementing self-reported data with other data-driven methods or studies. For example, studies that isolate interface-specific effects or examine how mixed-interface teams coordinate could provide a more precise view of how design differences shape practice.

\subsubsection{Understanding Moderators' Values During Report Review}

Our analysis has focused on how mods describe their use of the modqueue in practice, but it only indirectly surfaced the values that guide their work. Certain priorities like fairness, minimizing harm, or efficiency emerged in passing, yet we did not explicitly ask respondents to reflect on what they value most during report review. Future work could dive deeper into these questions, examining how moderators define their goals, how those goals shape the way they use the modqueue, and how different values and goals are in alignment, or in some cases, misaligned. For instance, one moderator may emphasize fairness by addressing oldest reports first, while another may focus on minimizing harm by tackling the most severe cases. Understanding these tradeoffs more explicitly would provide a clearer picture of how moderators navigate competing objectives, and could inform the design of tools and interventions that better align with the underlying objectives driving their work.

\subsubsection{Opportunities to Optimize Report Review Workflows}

Many moderators in our study described approaching report review in a systematic, checklist-like manner, working through items one at a time until the backlog was cleared. Future work could investigate how different approaches to item-by-item review influence both efficiency and quality of outcomes. This includes not only the strategies moderators already describe---such as prioritizing severe or time-sensitive reports, deferring ambiguous cases, or grouping similar items---but also the use of additional tools or interface features that support these strategies. Building a clearer understanding of how such approaches affect moderator performance could inform targeted improvements to existing workflows or motivate the development of new tools that better align with moderators’ needs. Such improvements would not replace the need for human judgment, but could reduce the routine effort involved in straightforward review and free up moderators’ attention for more complex or high-stakes cases.
\section{Conclusion}

By surveying 110 Reddit moderators across more than 400 subreddits, we examined how moderators use the platform’s modqueue and the challenges they face. We found that the modqueue is not a uniform or universally trusted workspace: some moderators treat it as a checklist for routine review, others repurposed it as a way to monitor community activity, and many leave the interface to gather additional context or coordinate with teammates. These findings contribute to a grounded understanding of the mechanics of report review and show how platform-provided interfaces both structure and constrain that process. In particular, we highlight that modqueue use is shaped by moderators’ goals, values, and community needs as much as by technical affordances. Building on this understanding, we contribute design implications that include supporting lightweight and configurable access to report context, re-framing the modqueue as a collaborative workspace rather than an item-by-item review list for individual moderators, and creating modular, cross-interface infrastructures that better reflect real-world workflows. These insights highlight opportunities to design or improve moderation interfaces so that they are more aligned with mods' preferred moderation workflows, supportive of moderators’ goals, and  reflective of the collaborative, value-driven nature of volunteer moderation work.

\appendix



\bibliographystyle{ACM-Reference-Format}
\bibliography{references}

\newpage


\end{document}